\documentclass[%
 reprint,
 amsmath,amssymb,
 aps,
]{revtex4-2}
\usepackage{graphicx}
\usepackage{dcolumn}
\usepackage{bm}
\begin{document}

\preprint{APS/123-QED}

\title{The enhanced ${\bf{s^*}}$-wave component of the superconducting gap in the overdoped HTSC cuprates with orthorhombic distortion}

\author{I.~A.~Makarov}
\email{maki@iph.krasn.ru}
\author{S.~G.~Ovchinnikov}%
 \email{sgo@iph.krasn.ru}
\affiliation{Kirensky Institute of Physics, Federal Research Center KSC SB RAS, 660036 Krasnoyarsk, Russia}
%
%

\date{\today}

\begin{abstract}
In this work, the concentration and temperature dependences of the superconducting gap of the HTSC cuprate in the orthorhombic phase are obtained within the framework of the Hubbard model taking into account the exchange mechanism of pairing. The enhanced $s^*$-wave component of the superconducting gap against the background of the prevailing $d$-wave component in a narrow range of hole concentrations in the overdoped region is found. To elucidate the reasons for such unusual behavior the electronic structure of the low-energy excitations and the structure of the contributions of the pair states to the superconducting gap in momentum space were investigated at different doping. It is shown that the presence of a shallow pocket in the region of the maximum value of the $s^*$-wave component of the gap results in different symmetry of the superconducting gap in different regions of the ${\bf{k}}$-space and an increase in the $s^*$-wave component. 
\end{abstract}

\maketitle


\section{\label{sec:Intro}Introduction}
The mechanism of superconducting pairing in HTSC cuprates is still a subject of discussion. The symmetry and magnitude of the superconducting gap are important characteristics that can indicate an active pairing mechanism. It is generally accepted that the superconducting gap in HTSC cuprates has the $d$-wave symmetry since a large number of experimental data of different types together supported this type of symmetry. Experiments on heat capacity~\cite{Moler94,Moler97,Momono94,Wright99}, penetration depth~\cite{Bonn92,Hardy93,Bonn96}, thermal conductivity~\cite{Salamon95,Yu96,Aubin97} indicate the presence of superconducting gap nodes. ARPES experiments that allow directly obtaining the ${\bf{k}}$-dependence of the gap have shown a strong anisotropy of the gap and a gap close to zero in the nodal direction~\cite{Shen93,Ding94,Shen95,Ding96}. Phase-sensitive experiments~\cite{Tsuei00,VanHarlingen95} demonstrated a shift of order $\pi$ in the phase of the order parameter between orthogonal directions $a$ and $b$ in YBa$_2$Cu$_3$O$_{7-\delta}$ (YBCO)~\cite{Wollman93,Wollman95,Brawner96_1,Brawner96_2,Mathai95} and half-integer flux quantum effect in YBCO~\cite{Tsuei94,Kirtley95}, Tl$_2$Ba$_2$CuO$_{4+\delta}$~\cite{Tsuei96} and Bi$_2$Sr$_2$CaCu$_2$O$_8$ (Bi2212)~\cite{Kirtley96}. All these results agree with what should be observed for the $d$-wave gap, but they do not exclude a slight admixture of components of different symmetry since the error and insufficient resolution do not allow us to assert a pure $d$-symmetry. For example, ARPES studies~\cite{Shen93,Ding94} directly admit the impossibility of distinguishing whether the superconducting gap in the nodal direction is zero or has a small but finite value. It is also necessary to bear in mind the sensitivity of experimental techniques to the surface and volume of the sample gap since the shape of the superconducting gap can differ at different depths in the sample~\cite{Muller02}. But the most important factors that can affect the symmetry of the superconducting gap are the structure phase and doping of the sample under study.

Among cuprates, there are compounds with both tetragonal and orthorhombic crystal lattice structures. Since it is believed that superconductivity is formed in CuO$_2$ layers then the gap should reflect the symmetry of the CuO$_2$ lattice. That is, real tetragonal and orthorhombic structures are reduced simply to square and rectangular lattices. The shape of the superconducting gap in real and reciprocal space refers to one of the irreducible representations of the symmetry group of the crystal lattice. The square lattice has four irreducible representations ${A_{1g}}$, ${A_{2g}}$, ${B_{1g}}$, ${B_{2g}}$, the corresponding states with singlet pairing are designated as $s+$, $g$, ${d_{{x^2} - {y^2}}}$, ${d_{xy}}$~\cite{Tsuei00,Annett96}. In the case of a rectangular lattice, the paired states $s+$ and ${d_{{x^2} - {y^2}}}$ belong to the same irreducible representation ${A_{1g}}$, that is, the superconducting gap is a mixture of $d$- and $s$-wave gaps~\cite{Sigrist87,Mineev98}. The superconducting gap cannot have pure $d$-wave symmetry in orthorhombic samples although the $s$-wave component is apparently very small since it does not manifest itself clearly in the nodal direction in ARPES experiments near optimal doping.

The form of the superconducting gap in ${\bf{k}}$-space reflects not just the ${\bf{k}}$-dependence of the pairing interaction, but also the joint ${\bf{k}}$-dependence of the energy, and the spectral weight of the electronic states. Doping significantly reconstructs the electronic structure of cuprates. Doping the parent compound in the state of an antiferromagnetic insulator leads to the destruction of the long-range magnetic order and leaves the short-range magnetic order. An unusual pseudogap electronic state appears in the region of weak and optimal doping. The pseudogap closes in the directions from the nodal to the anti-nodal with increasing doping. At a certain concentration of doped holes $p^*$ the pseudogap disappears over the entire Fermi surface and the electrons begin to behave like a normal paramagnetic Fermi liquid while the superconducting phase still exists.

Since the electronic structure of cuprates changes greatly with doping it is logical to expect that the ${\bf{k}}$-dependence of the superconducting gap will also change depending on the number of charge carriers even if the pairing interaction does not depend on the carrier concentration. Indeed, it was shown in~\cite{Kelley96} that the degree of anisotropy of the superconducting gap changes with increasing doping. The opening of the gap in the nodal direction was found with oxygen doping when passing from underdoped to the overdoped sample of Bi2212, indicating the absence of pure $d$-wave symmetry of the gap in the region of strong doping~\cite{Kelley96}. The change in the size of the superconducting gap in the nodal direction with doping was obtained in ARPES on Bi2212 in~\cite{Zhao01}. While in a slightly overdoped compound the gap in the nodal direction had a small value which can be considered as zero for the $d$-wave order parameter, the nodal gap is significant in a heavily overdoped compound and cannot be considered as zero. The results of these and other experiments can be interpreted in terms of a two-component order parameter. The theoretical works~\cite{Kotliar88,Li93,Betouras95} support the conclusion about the presence of the two-component order parameter in cuprates.

ARPES in overdoped YBCO~\cite{Lu01} shows the difference in the superconducting gap magnitude and superconducting peak intensity along the $x$ and $y$ axes. These features indicate deviation from the $d$-wave pairing state. Raman scattering spectra also demonstrate $x - y$ difference~\cite{Limonov00}. Energies of pair-breaking peaks of the imaginary part of electronic response function in $xx$ and $yy$ polarizations and their peak widths are different in YBCO. These features are most pronounced in overdoped compound although are also observed in underdoped and optimally doped samples. The position of the pair-breaking peak in the Raman spectra serves as an indicator of the gap size, and the relative position of these peaks in lines of different symmetry indicates its symmetry. The specific features of strong doping obtained from Raman spectra are a decrease in the pair-breaking peaks energy and change in the ratio of pair-breaking peaks in the spectra of different symmetry with increasing doping, change in the intensity of pair-breaking peaks, which has a different character for spectra of different symmetry~\cite{Friedl90,Devereaux95,Hackl96,Strohm97,Nemetschek98,Masui03,Hiramachi07}. Raman spectra in optimally and overdoped cuprates YBCO, Bi2212, and Hg1212 and the effect of the decrease in the polarization dependence of the gap energy were explained in~\cite{Nemetschek98} by considering the $d+s$ symmetry of the gap.

The form of the dependences of the tunnel conductance on the voltage in scanning tunneling spectroscopy in underdoped and optimally doped YBCO~\cite{Yeh01} are consistent with theories for the $d$-symmetry of the gap. However, the overdoped Ca-doped YBCO exhibited symmetric subgap peaks which indicate doping-induced variations in the pairing symmetry~\cite{Yeh01}. The spectra of overdoped compounds were described using $d+s$ symmetry~\cite{Yeh01}.

Evidence of the $s$-wave component of the superconducting gap in orthorhombic cuprates was observed in Josephson tunneling~\cite{Sun94,Kleiner96}, in $a-b$ anisotropy measurements of the penetration depth in the microwave and far-infrared spectroscopic measurements~\cite{Zhang94,Basov95}. The gap with $d$- and $s$-wave components was found in measurements of thermal conductivity for YBCO~\cite{Aubin97}. Two peaks in the density of states which are characteristic of the orthorhombic phase~\cite{Bealmonod96} were observed in vortex imaging experiments~\cite{Maggioaprile95}.

Theoretical considerations~\cite{Eremin95,Plakida00} also demonstrate admixing of $s^*$-wave symmetry to the $d$-wave gap in the orthorhombic phase of cuprates. A large proportion of the $s^*$-wave component was obtained in the work~\cite{Plakida00} within the $t-J$ model but its concentration dependence is not discussed. The dependence of $d$- and $s^*$-wave components of the superconducting gap ratio on the chemical potential level was found in work~\cite{Eremin95}. The variation in this ratio is the main result of doping within the rigid band model, but the joint change of the electronic structure and the ratio of the components with doping have not been investigated.

It can be seen from the above experimental data that the $s$-wave component is admixed to the $d$-wave component of the superconducting gap in the region of strong doping. At least two possible reasons for this behavior with doping can be indicated: a change in the pairing mechanism, or a change in the electronic structure. Since it is known that the change of carrier concentration from optimal to strong doping results in the transformation of the electronic system from the pseudogap state to the normal Fermi liquid, the second possibility seems to be more realistic. In this paper, we theoretically investigate the dependence of the superconducting gap on doping and temperature and try to understand the mechanism of the change in the ratio of the superconducting gap components with doping. The theory of superconductivity of the BCS type is constructed within the framework of the Hubbard model for excitations in a plane rectangular lattice. The system of equations for the $d$- and the extended $s^*$-wave components of the superconducting gap is obtained taking into account the anisotropic exchange mechanism of pairing. The self-consistent solution of this system makes it possible to obtain gaps for different doping values, temperatures, and pairing constants. A significant change in the ratio of the $d$- and the $s^*$-wave components with doping was actually found in this approach. The dispersion surface and the Fermi contour were calculated to determine the structure of the electronic states whose pairing forms the superconducting gap. Further analysis of the ${\bf{k}}$-dependent contributions of these states to the components of the superconducting gap made it possible to find out which paired states are responsible for the observed features of the concentration dependence of the superconducting gap.

The paper includes six sections. Section~\ref{sec:method} describes the Hubbard model Hamiltonian of excitations in the rectangular lattice and reports the method for calculating the superconducting gap. Section~\ref{sec:dsgaps} contains the concentration and temperature dependences of $d$- and $s^*$-wave components of the superconducting gap. Section~\ref{sec:elstr} reports the electronic structure of low-energy excitations in the normal and superconducting phases and their transformation with doping in the overdoped region. In section~\ref{sec:partcontr} we will analyze contributions of electronic states with different wave vectors to $d$- and $s^*$-wave components of the superconducting gap. In Section~\ref{sec:Concl} the main results are summarized.

\section{\label{sec:method}Superconducting state in the Hubbard model with ${\bf{D}}$- and extended ${\bf{S^{\ast}}}$-wave gap components}
There is a large number of theoretical papers devoted to superconductivity in the Hubbard model in 2d square lattice, see recent reviews~\cite{Rohringer18,Kresin21}, and a few papers concerning the orthorhombic phase. The description of the normal phase and some properties of the superconducting phase of HTSC cuprates in the framework of a realistic $p-d$ model for a layer of CuO$_6$ octahedra with orthorhombic distortion was carried out in~\cite{Makarov21}. In the regime of strong electron correlations, the Hubbard model appears as the low energy effective model within the Hubbard operators representation with parameters obtained by the generalized tight-binding (GTB) method~\cite{Ovchinnikov89,Gavrichkov00,Korshunov05}. In this work, we will use this model to calculate the electronic structure of the quasiparticle excitations in the rectangular lattice with $N$ sites in the superconducting phase.

The basis of the Hubbard model consists of four local eigenstates $\left| 0 \right\rangle$, $\left| \sigma  \right\rangle$, $\left| S \right\rangle $ with $\sigma  =  \pm {1 \mathord{\left/
 {\vphantom {1 2}} \right.
 \kern-\nulldelimiterspace} 2}$ for each site of the lattice. Hamiltonian of the Hubbard model within the Hubbard X-operators is given by:
\begin{eqnarray}
H && = \sum\limits_f {{\varepsilon _0}X_f^{00}}  + \sum\limits_{f\sigma } {\left( {{\varepsilon _\sigma } - \mu } \right)X_f^{\sigma \sigma }}  + \\\nonumber
&&\sum\limits_f {\left( {{\varepsilon _S} - 2\mu } \right)X_f^{SS}}  + \sum\limits_{fg\sigma } {\sum\limits_{pqmn} {{t_{fg\sigma }}\left( {pq,mn} \right)X_f^{pq}X_g^{mn}} },
\label{eq:Ham} 
\end{eqnarray}
where ${\varepsilon _0}$, ${\varepsilon _\sigma }$, ${\varepsilon _S}$ are energies of states $\left| 0 \right\rangle$, $\left| \sigma  \right\rangle$, $\left| S \right\rangle $, respectively, $\mu $ is the chemical potential. ${t_{fg\sigma }}\left( {pq,mn} \right)$ are hopping integrals between quasiparticle excitations $\left( {pq} \right)$ and $\left( {mn} \right)$ in the sites $f$ and $g$, respectively. There are four different Fermi-type quasiparticles, corresponding to electron removal and electron addition excitations $\left\{ {\left( {0\bar \sigma } \right),\left( {\sigma S} \right),\left( {\sigma 0} \right),\left( {S\bar \sigma } \right)} \right\}$. We will consider one value of the orthorhombicity factor for the CuO$_2$ lattice: ${{\left( {b - a} \right)} \mathord{\left/
 {\vphantom {{\left( {b - a} \right)} {a = 4.15}}} \right.
 \kern-\nulldelimiterspace} {a = 4.15}}\% $, where $a$, $b$ are lattice parameters. We take the numerical values of the Hamiltonian parameters ${\varepsilon _0}$, ${\varepsilon _\sigma }$, ${\varepsilon _S}$, ${t_{fg\sigma }}\left( {pq,mn} \right)$ from the work~\cite{Makarov21}.
 
 In this work, we use the same technique to obtain a spectrum of quasiparticles and a superconducting gap for a system with orthorhombic distortion as used in~\cite{Plakida03} for describing the tetragonal phase. The spectrum of quasiparticles and the superconducting gap were obtained using the method of equations of motion for Green's function. Green's function based on basis of Hubbard model is $4\times4$ matrix (which consists of normal and anomalous $2\times2$ matrix):
\begin{widetext}
\begin{eqnarray}
{\hat D^{\bar \sigma }}\left( {f,g} \right) = \left[ {\begin{array}{*{20}{c}}
{\left\langle {\left\langle {{X_f^{0\bar \sigma }}}
 \mathrel{\left | {\vphantom {{X_f^{0\bar \sigma }} {X_g^{\bar \sigma 0}}}}
 \right. \kern-\nulldelimiterspace}
 {{X_g^{\bar \sigma 0}}} \right\rangle } \right\rangle }&{\left\langle {\left\langle {{X_f^{0\bar \sigma }}}
 \mathrel{\left | {\vphantom {{X_f^{0\bar \sigma }} {X_g^{S\sigma }}}}
 \right. \kern-\nulldelimiterspace}
 {{X_g^{S\sigma }}} \right\rangle } \right\rangle }&{\left\langle {\left\langle {{X_f^{0\bar \sigma }}}
 \mathrel{\left | {\vphantom {{X_f^{0\bar \sigma }} {X_g^{0\sigma }}}}
 \right. \kern-\nulldelimiterspace}
 {{X_g^{0\sigma }}} \right\rangle } \right\rangle }&{\left\langle {\left\langle {{X_f^{0\bar \sigma }}}
 \mathrel{\left | {\vphantom {{X_f^{0\bar \sigma }} {X_g^{\bar \sigma S}}}}
 \right. \kern-\nulldelimiterspace}
 {{X_g^{\bar \sigma S}}} \right\rangle } \right\rangle }\\
{\left\langle {\left\langle {{X_f^{\sigma S}}}
 \mathrel{\left | {\vphantom {{X_f^{\sigma S}} {X_g^{\bar \sigma 0}}}}
 \right. \kern-\nulldelimiterspace}
 {{X_g^{\bar \sigma 0}}} \right\rangle } \right\rangle }&{\left\langle {\left\langle {{X_f^{\sigma S}}}
 \mathrel{\left | {\vphantom {{X_f^{\sigma S}} {X_g^{S\sigma }}}}
 \right. \kern-\nulldelimiterspace}
 {{X_g^{S\sigma }}} \right\rangle } \right\rangle }&{\left\langle {\left\langle {{X_f^{\sigma S}}}
 \mathrel{\left | {\vphantom {{X_f^{\sigma S}} {X_g^{0\sigma }}}}
 \right. \kern-\nulldelimiterspace}
 {{X_g^{0\sigma }}} \right\rangle } \right\rangle }&{\left\langle {\left\langle {{X_f^{\sigma S}}}
 \mathrel{\left | {\vphantom {{X_f^{\sigma S}} {X_g^{\bar \sigma S}}}}
 \right. \kern-\nulldelimiterspace}
 {{X_g^{\bar \sigma S}}} \right\rangle } \right\rangle }\\
{\left\langle {\left\langle {{X_f^{\sigma 0}}}
 \mathrel{\left | {\vphantom {{X_f^{\sigma 0}} {X_g^{\bar \sigma 0}}}}
 \right. \kern-\nulldelimiterspace}
 {{X_g^{\bar \sigma 0}}} \right\rangle } \right\rangle }&{\left\langle {\left\langle {{X_f^{\sigma 0}}}
 \mathrel{\left | {\vphantom {{X_f^{\sigma 0}} {X_g^{S\sigma }}}}
 \right. \kern-\nulldelimiterspace}
 {{X_g^{S\sigma }}} \right\rangle } \right\rangle }&{\left\langle {\left\langle {{X_f^{\sigma 0}}}
 \mathrel{\left | {\vphantom {{X_f^{\sigma 0}} {X_g^{0\sigma }}}}
 \right. \kern-\nulldelimiterspace}
 {{X_g^{0\sigma }}} \right\rangle } \right\rangle }&{\left\langle {\left\langle {{X_f^{\sigma 0}}}
 \mathrel{\left | {\vphantom {{X_f^{\sigma 0}} {X_g^{\bar \sigma S}}}}
 \right. \kern-\nulldelimiterspace}
 {{X_g^{\bar \sigma S}}} \right\rangle } \right\rangle }\\
{\left\langle {\left\langle {{X_f^{S\bar \sigma }}}
 \mathrel{\left | {\vphantom {{X_f^{S\bar \sigma }} {X_g^{\bar \sigma 0}}}}
 \right. \kern-\nulldelimiterspace}
 {{X_g^{\bar \sigma 0}}} \right\rangle } \right\rangle }&{\left\langle {\left\langle {{X_f^{S\bar \sigma }}}
 \mathrel{\left | {\vphantom {{X_f^{S\bar \sigma }} {X_g^{S\sigma }}}}
 \right. \kern-\nulldelimiterspace}
 {{X_g^{S\sigma }}} \right\rangle } \right\rangle }&{\left\langle {\left\langle {{X_f^{S\bar \sigma }}}
 \mathrel{\left | {\vphantom {{X_f^{S\bar \sigma }} {X_g^{0\sigma }}}}
 \right. \kern-\nulldelimiterspace}
 {{X_g^{0\sigma }}} \right\rangle } \right\rangle }&{\left\langle {\left\langle {{X_f^{S\bar \sigma }}}
 \mathrel{\left | {\vphantom {{X_f^{S\bar \sigma }} {X_g^{\bar \sigma S}}}}
 \right. \kern-\nulldelimiterspace}
 {{X_g^{\bar \sigma S}}} \right\rangle } \right\rangle }
\end{array}} \right].
\label{eq:matrD}
\end{eqnarray}
\end{widetext}
The general form of the system of the equations of motion for Fourier transform of each component of matrix Green's function ${\hat D^{\bar \sigma }}\left( {f,g} \right)$ has a view:
\begin{eqnarray}
\label{eq:eqmot}
\omega D_{\left( {pq} \right)\left( {q'p'} \right)}^{\bar \sigma }\left( {{\bf{k}};\omega } \right)& = &{\delta _{pp'}}{\delta _{qq'}}F\left( {pq} \right) + \\
&&\Omega \left( {pq} \right)D_{\left( {pq} \right)\left( {q'p'} \right)}^{\bar \sigma }\left( {{\bf{k}};\omega } \right) + L_j^{pq},\nonumber
\end{eqnarray}
where operators $L_j^{pq}$ include higher-order Green's functions. $\Omega \left( m \right) = \Omega \left( {pq} \right) = {\varepsilon _p} - {\varepsilon _q} - \mu $ is the energy of the quasiparticle $\left( {pq} \right)$. $F\left( {pq} \right) = \left\langle {{X^{pp}}} \right\rangle  + \left\langle {{X^{qq}}} \right\rangle $ is the filling factor. The filling numbers of states $\left\langle {{X^{pp}}} \right\rangle $ are calculated self-consistently with chemical potential and kinematic correlators (Appendix~\ref{app:fillnumb}).

System of equations (\ref{eq:eqmot}) for all components of matrix Green's function ${\hat D^{\bar \sigma }}\left( {{\bf{k}};\omega } \right)$ is decoupled within generalized mean-field approximation using projection technique of Mori-Zwanzig formalism type~\cite{Plakida03}. In this decoupling method, operators $L_j^{pq}$ are represented as a sum of reducible and irreducible parts. Reducible part can be linearized over X-operators of Hubbard model basis:
\begin{widetext}
\begin{eqnarray}
L_j^{0\bar \sigma } & = & \sum\limits_{lpq} {T_j^{\left( {0\bar \sigma } \right)(qp)}{\rm X}_l^{(pq)}}  + L_j^{\left( {0\bar \sigma } \right)\left( {irr} \right)} \to \\\nonumber
&& \to \sum\limits_l {\left( {T_j^{\left( {0\bar \sigma } \right)(\bar \sigma 0)}{\rm X}_l^{(0\bar \sigma )} + T_j^{\left( {0\bar \sigma } \right)(S\sigma )}{\rm X}_l^{(\sigma S)} + \Delta _j^{\left( {0\bar \sigma } \right)(0\sigma )}{\rm X}_l^{(\sigma 0)} + \Delta _j^{\left( {0\bar \sigma } \right)(\bar \sigma S)}{\rm X}_l^{(S\bar \sigma )}} \right)}  + L_j^{\left( {0\bar \sigma } \right)\left( {irr} \right)},
\label{eq:irrL}
\end{eqnarray}
\end{widetext}
where $T_j^{\left( {0\bar \sigma } \right)\left( {qp} \right)} = {{\left\langle {\left\{ {L_j^{0\bar \sigma },X_l^{qp}} \right\}} \right\rangle } \mathord{\left/
 {\vphantom {{\left\langle {\left\{ {L_j^{0\bar \sigma },X_l^{qp}} \right\}} \right\rangle } {\left\langle {\left\{ {{\rm X}_l^{pq},X_l^{qp}} \right\}} \right\rangle }}} \right.
 \kern-\nulldelimiterspace} {\left\langle {\left\{ {{\rm X}_l^{pq},X_l^{qp}} \right\}} \right\rangle }}$, $\Delta _j^{\left( {0\bar \sigma } \right)\left( {pq} \right)} = {{\left\langle {\left\{ {L_j^{0\bar \sigma },X_l^{pq}} \right\}} \right\rangle } \mathord{\left/
 {\vphantom {{\left\langle {\left\{ {L_j^{0\bar \sigma },X_l^{pq}} \right\}} \right\rangle } {\left\langle {\left\{ {{\rm X}_l^{qp},X_l^{pq}} \right\}} \right\rangle }}} \right.
 \kern-\nulldelimiterspace} {\left\langle {\left\{ {{\rm X}_l^{qp},X_l^{pq}} \right\}} \right\rangle }}$ are coefficients of linearization. Terms $T_j^{\left( {pq} \right)\left( {q'p'} \right)}$ define the self-energy operator, in generalized mean-field approximation they contain hoppings, kinematic, and spin-spin correlation functions. Terms $\Delta _{\bf{k}}^{\alpha \beta }$ are superconducting gap functions. Definitions of terms $T_{\bf{k}}^{\alpha \beta }$ are given in Appendix~\ref{app:kinpar}. Further, in the generalized mean-field approximation we neglect the irreducible operator $L_j^{pq\left( {irr} \right)}$. The Green's function ${\hat D^{\bar \sigma }}\left( {{\bf{k}};\omega } \right)$ is defined by the expression:
\begin{equation}
{\hat D^{\bar \sigma }}\left( {{\bf{k}};\omega } \right) = {\hat R^{ - 1}}\left( {{\bf{k}};\omega } \right)\hat F,
\label{eq:Dyson}
\end{equation}
where
\begin{widetext}
\begin{eqnarray}
{\hat R_{\bf{k}}} = \left[ {\begin{array}{*{20}{c}}
{\omega  - \Omega \left( {0\bar \sigma } \right) - \bar T_{\bf{k}}^{11}}&{ - \bar T_{\bf{k}}^{12}}&{ - \Delta _{\bf{k}}^{11}}&{ - \Delta _{\bf{k}}^{12}}\\
{ - \bar T_{\bf{k}}^{21}}&{\omega  - \Omega \left( {\sigma S} \right) - \bar T_{\bf{k}}^{22}}&{ - \Delta _{\bf{k}}^{21}}&{ - \Delta _{\bf{k}}^{22}}\\
{ - \Delta {{_{\bf{k}}^{11}}^ * }}&{\Delta {{_{\bf{k}}^{12}}^ * }}&{\omega  + \Omega \left( {0\sigma } \right) + T_{\bf{k}}^{11}}&{T_{\bf{k}}^{21}}\\
{\Delta {{_{\bf{k}}^{21}}^ * }}&{ - \Delta {{_{\bf{k}}^{22}}^ * }}&{T_{\bf{k}}^{12}}&{\omega  + \Omega \left( {\bar \sigma S} \right) + T_{\bf{k}}^{22}}
\end{array}} \right]
\label{eq:enDyson}
\end{eqnarray}
\end{widetext}
and $\hat F$ is the matrix of filling factors
\begin{eqnarray}
\hat F = \left[ {\begin{array}{*{20}{c}}
{F\left( {0\sigma } \right)}&0&0&0\\
0&{F\left( {\bar \sigma S} \right)}&0&0\\
0&0&{F\left( {\bar \sigma 0} \right)}&0\\
0&0&0&{F\left( {S\sigma } \right)}
\end{array}} \right].
\label{eq:Ffact}
\end{eqnarray}
The electronic structure of the normal phase with orthorhombic symmetry has been studied within the same approach in the paper~\cite{Makarov21}. In undoped cuprate, it has the empty upper Hubbard subband (UHB) and the occupied lower Hubbard subband (LHB) with the insulator gap. To be closer to the general language of semiconductors and insulators, we use standard terminology of a conductivity band (index 1) and a valence band 2. With hole doping the Fermi level moves to the top of the valence band with quite complicate changes of the Fermi surface topology (Lifshitz transitions), discussed in~\cite{Makarov21}. 

Superconducting gaps for the pairing of quasiparticles inside conductivity band $\Delta _{\bf{k}}^{11}$, inside valence band $\Delta _{\bf{k}}^{22}$, and between these bands $\Delta _{\bf{k}}^{12}$, $\Delta _{\bf{k}}^{21}$ are:
\begin{eqnarray}
\label{eq:gaps}
\Delta _{\bf{k}}^{11} & = &  - \frac{1}{{NF\left( {0\sigma } \right)}}\sum\limits_{\bf{q}} {\frac{4}{{{E_{ct}}}}{{\left( {{{\left( {{t^{12}}} \right)}^2}} \right)}_{{\bf{k}} - {\bf{q}}}}{{\left\langle {{X^{\sigma S}}{X^{\bar \sigma S}}} \right\rangle }_{\bf{q}}}}, \\\nonumber
\Delta _{\bf{k}}^{21} & = & \frac{{\left( { - 1} \right)}}{{NF\left( {0\sigma } \right)}}\sum\limits_{\bf{q}} {\frac{{2{{\left( {\left( {{t^{11}} + {{\bar t}^{22}}} \right){t^{12}}} \right)}_{{\bf{k}} - {\bf{q}}}}}}{{{E_{ct}}}}{{\left\langle {{X^{\sigma S}}{X^{\bar \sigma S}}} \right\rangle }_{\bf{q}}}}, \\\nonumber 
\Delta _{\bf{k}}^{12} & = & \frac{1}{{NF\left( {\bar \sigma S} \right)}}\sum\limits_{\bf{q}} {\frac{{2{{\left( {{t^{12}}\left( {{{\bar t}^{11}} + {t^{22}}} \right)} \right)}_{{\bf{k}} - {\bf{q}}}}}}{{{E_{ct}}}}{{\left\langle {{X^{\sigma S}}{X^{\bar \sigma S}}} \right\rangle }_{\bf{q}}}}, \\\nonumber
\Delta _{\bf{k}}^{22} & = &  - \frac{1}{{NF\left( {\bar \sigma S} \right)}}\sum\limits_{\bf{q}} {\frac{4}{{{E_{ct}}}}{{\left( {{{\left( {{t^{12}}} \right)}^2}} \right)}_{{\bf{k}} - {\bf{q}}}}{{\left\langle {{X^{\sigma S}}{X^{\bar \sigma S}}} \right\rangle }_{\bf{q}}}}.
\end{eqnarray}
The terms ${t^{\alpha \beta }}$ are the same intraband and interband hopping integrals as ${t}\left( {pq,mn} \right)$ but written in a more compact form using indices $1$ and $2$; their more detailed description is given in Appendix~\ref{app:kinpar}. The superconducting pairing in each of the four Eqs.~(\ref{eq:gaps}) results from the exchange mechanism. For hole-doped cuprates, we take into account pairing only inside the valence band. Therefore, in the rest of the work, we will talk about two components of the superconducting gap $\Delta _{\bf{k}}^{22}$ which will be denoted simply as ${\Delta _{\bf{k}}}$. Then the superconducting gap can be written as
\begin{equation}
{\Delta _{\bf{k}}} = \frac{1}{N}\sum\limits_{\bf{q}} {\frac{4}{{{E_{ct}}}}t_{{\bf{k}} - {\bf{q}}}^2\left( {\bar \sigma 0,\sigma S} \right){{\left\langle {{X^{\sigma S}}{X^{\bar \sigma S}}} \right\rangle }_{\bf{q}}}},
\label{eq:gapeq}
\end{equation}
where ${{t_{{\bf{k}} - {\bf{q}}}^2\left( {\bar \sigma 0,\sigma S} \right)} \mathord{\left/
 {\vphantom {{t_{{\bf{k}} - {\bf{q}}}^2\left( {\bar \sigma 0,\sigma S} \right)} {{E_{ct}}}}} \right.
 \kern-\nulldelimiterspace} {{E_{ct}}}} = {J_{{\bf{k}} - {\bf{q}}}}$ is the effective exchange interaction similar to antiferromagnetic superexchange interaction in the $t-J$ model~\cite{Anderson63,Plakida03}, ${E_{ct}} = \Omega \left( {\sigma S} \right) - \Omega \left( {0\bar \sigma } \right)$ is the charge transfer gap between dispersionless upper and lower Hubbard bands, which is similar to the effective Hubbard ${U_{eff}}$. An anomalous average ${\left\langle {{X^{\sigma S}}{X^{\bar \sigma S}}} \right\rangle _{\bf{q}}}$ can be obtained from Green's function $\left\langle {\left\langle {{X_{\bf{q}}^{\bar \sigma S}}}
 \mathrel{\left | {\vphantom {{X_{\bf{q}}^{\bar \sigma S}} {X_{\bf{q}}^{\sigma S}}}}
 \right. \kern-\nulldelimiterspace}
 {{X_{\bf{q}}^{\sigma S}}} \right\rangle } \right\rangle $. Since Green's function $\left\langle {\left\langle {{X_{\bf{q}}^{\bar \sigma S}}}
 \mathrel{\left | {\vphantom {{X_{\bf{q}}^{\bar \sigma S}} {X_{\bf{q}}^{\sigma S}}}}
 \right. \kern-\nulldelimiterspace}
 {{X_{\bf{q}}^{\sigma S}}} \right\rangle } \right\rangle$ depends on the superconducting gap we obtain a self-consistent equation:
\begin{equation}
{\Delta _{\bf{k}}} = \frac{1}{N}\sum\limits_{\bf{q}} {{J_{{\bf{k}} - {\bf{q}}}}{\Delta _{\bf{q}}}{A_{\bf{q}}}},
\label{eq:gapeq_short}
\end{equation}
where
\begin{widetext}
\begin{eqnarray}
{A_{\bf{q}}}& = &\frac{1}{{{B_{\bf{q}}}}}\left[ {\frac{{\varepsilon _{1{\bf{q}}}^2 - \xi _{1{\bf{q}}}^2}}{{{\varepsilon _{1{\bf{q}}}}}}th\frac{{{\varepsilon _{1{\bf{q}}}}}}{{2kT}} - \frac{{\varepsilon _{2{\bf{q}}}^2 - \xi _{1{\bf{q}}}^2}}{{{\varepsilon _{2{\bf{q}}}}}}th\frac{{{\varepsilon _{2{\bf{q}}}}}}{{2kT}}} \right], \\\nonumber
{B_{\bf{q}}}& = &\frac{1}{2}\sqrt {{{\left( {\xi _{1{\bf{q}}}^2 + \xi _{2{\bf{q}}}^2 + 2\bar T_{\bf{q}}^{12}\bar T_{\bf{q}}^{21} + {{\left| {{\Delta _{\bf{q}}}} \right|}^2}} \right)}^2} - 4\left( {{{\left( {{\xi _{1{\bf{q}}}}{\xi _{1{\bf{q}}}} - \bar T_{\bf{q}}^{12}\bar T_{\bf{q}}^{21}} \right)}^2} + \xi _{1{\bf{q}}}^2{{\left| {{\Delta _{\bf{q}}}} \right|}^2}} \right)},\nonumber
\end{eqnarray}
\end{widetext} 
${\xi _{1{\bf{q}}}} = \Omega \left( {0\bar \sigma } \right) + \bar T_{\bf{q}}^{11}$, ${\xi _{2{\bf{q}}}} = \Omega \left( {\sigma S} \right) + \bar T_{\bf{q}}^{22}$ are the dispersions of quasiparticle excitations inside conductivity (valence) band without interband hybridization in the normal phase, $ \pm {\varepsilon _{1{\bf{q}}}}, \pm {\varepsilon _{2{\bf{q}}}}$ are the dispersions of the Bogolyubov quasiparticle bands, ${\varepsilon _{1{\bf{q}}}} = \frac{1}{2}\sqrt {{{\left( {\xi _{1{\bf{q}}}^2 + \xi _{2{\bf{q}}}^2 + 2\bar T_{\bf{q}}^{12}\bar T_{\bf{q}}^{21} + {{\left| {{\Delta _{\bf{q}}}} \right|}^2}} \right)}^2} + {B_{\bf{q}}}} $, ${\varepsilon _{2{\bf{q}}}} = \frac{1}{2}\sqrt {{{\left( {\xi _{1{\bf{q}}}^2 + \xi _{2{\bf{q}}}^2 + 2\bar T_{\bf{q}}^{12}\bar T_{\bf{q}}^{21} + {{\left| {{\Delta _{\bf{q}}}} \right|}^2}} \right)}^2} - {B_{\bf{q}}}} $. In the orthorhombic phase, the superconducting gap has a form of the sum of $d$-wave symmetry and extended $s^*$-wave symmetry: ${\Delta _{\bf{k}}} = {\Delta _d}\left( {\cos \left( {{k_x}} \right) - \cos \left( {{k_y}} \right)} \right) + {\Delta _{s*}}\left( {\cos \left( {{k_x}} \right) + \cos \left( {{k_y}} \right)} \right)$. Equating the coefficients of independent functions $\cos \left( {{k_x}} \right)$ and $\cos \left( {{k_y}} \right)$ in the left and right sides of the Eq.~(\ref{eq:gapeq_short}) we obtain a system of equations for linear combinations of the superconducting gap components in the approximation of nearest-neighbor hopping:
\begin{widetext}
\begin{eqnarray}
\label{eq:gapdssum}
\left\{ {\begin{array}{*{20}{c}}
{{\Delta _d} + {\Delta _{s*}} = \frac{1}{N}\sum\limits_{\bf{q}} {2J_{01}^x{A_{\bf{q}}}\left[ {{{\cos }^2}\left( {{q_x}} \right)\left( {{\Delta _d} + {\Delta _{s*}}} \right) + \cos \left( {{q_x}} \right)\cos \left( {{q_y}} \right)\left( {{\Delta _{s*}} - {\Delta _d}} \right)} \right]} },\\
{{\Delta _{s*}} - {\Delta _d} = \frac{1}{N}\sum\limits_{\bf{q}} {2J_{01}^y{A_{\bf{q}}}\left[ {\cos \left( {{q_x}} \right)\cos \left( {{q_y}} \right)\left( {{\Delta _d} + {\Delta _{s*}}} \right) + {{\cos }^2}\left( {{q_y}} \right)\left( {{\Delta _{s*}} - {\Delta _d}} \right)} \right]} }.
\end{array}} \right.
\end{eqnarray}
\end{widetext}
$J_{01}^x$ and $J_{01}^y$ are exchange interaction parameters between nearest neighbors along the $x$ and $y$ axis, respectively, $J_{01}^x = {\rm{0}}{\rm{.168}}$ eV and $J_{01}^y = 0.164$ eV at the orthorhombic distortion ${{\delta b} \mathord{\left/
 {\vphantom {{\delta b} a}} \right.
 \kern-\nulldelimiterspace} a} = 4.15\% $~\cite{Makarov21}. The system of Eqs.~(\ref{eq:gapdssum}) for components $\left\{ {{\Delta _d},{\Delta _{s*}}} \right\}$ has the form:
\begin{widetext}
\begin{eqnarray}
{\Delta _d} = \frac{1}{N}\sum\limits_{\bf{q}} {\left[ {{A_{\bf{q}}}\left( {J_{01}^x{{\cos }^2}\left( {{q_x}} \right) - \left( {J_{01}^x + J_{01}^y} \right)\cos \left( {{q_x}} \right)\cos \left( {{q_y}} \right) + J_{01}^y{{\cos }^2}\left( {{q_y}} \right)} \right){\Delta _d} + } \right.} \\\nonumber
\left. {{A_{\bf{q}}}\left( {J_{01}^x{{\cos }^2}\left( {{q_x}} \right) + \left( {J_{01}^x - J_{01}^y} \right)\cos \left( {{q_x}} \right)\cos \left( {{q_y}} \right) - J_{01}^y{{\cos }^2}\left( {{q_y}} \right)} \right){\Delta _{s*}}} \right],\\\nonumber
{\Delta _{s*}} = \frac{1}{N}\sum\limits_{\bf{q}} {\left[ {{A_{\bf{q}}}\left( {J_{01}^x{{\cos }^2}\left( {{q_x}} \right) - \left( {J_{01}^x - J_{01}^y} \right)\cos \left( {{q_x}} \right)\cos \left( {{q_y}} \right) - J_{01}^y{{\cos }^2}\left( {{q_y}} \right)} \right){\Delta _d} + } \right.} \\\nonumber
\left. {{A_{\bf{q}}}\left( {J_{01}^x{{\cos }^2}\left( {{q_x}} \right) + \left( {J_{01}^x + J_{01}^y} \right)\cos \left( {{q_x}} \right)\cos \left( {{q_y}} \right) + J_{01}^y{{\cos }^2}\left( {{q_y}} \right)} \right){\Delta _{s*}}} \right].
\label{eq:gap_d_s_selfc}
\end{eqnarray}
\end{widetext}.
Introducing the designations
\begin{widetext}
\begin{eqnarray}
{D_{1{\bf{q}}}} = \frac{{{A_{\bf{q}}}}}{N}\left( {J_{01}^x{{\cos }^2}\left( {{q_x}} \right) - \left( {J_{01}^x + J_{01}^y} \right)\cos \left( {{q_x}} \right)\cos \left( {{q_y}} \right) + J_{01}^y{{\cos }^2}\left( {{q_y}} \right)} \right),\\\nonumber
{D_{2{\bf{q}}}} = \frac{{{A_{\bf{q}}}}}{N}\left( {J_{01}^x{{\cos }^2}\left( {{q_x}} \right) + \left( {J_{01}^x - J_{01}^y} \right)\cos \left( {{q_x}} \right)\cos \left( {{q_y}} \right) - J_{01}^y{{\cos }^2}\left( {{q_y}} \right)} \right),\\\nonumber
{S_{1{\bf{q}}}} = \frac{{{A_{\bf{q}}}}}{N}\left( {J_{01}^x{{\cos }^2}\left( {{q_x}} \right) - \left( {J_{01}^x - J_{01}^y} \right)\cos \left( {{q_x}} \right)\cos \left( {{q_y}} \right) - J_{01}^y{{\cos }^2}\left( {{q_y}} \right)} \right),\\\nonumber
{S_{2{\bf{q}}}} = \frac{{{A_{\bf{q}}}}}{N}\left( {J_{01}^x{{\cos }^2}\left( {{q_x}} \right) + \left( {J_{01}^x + J_{01}^y} \right)\cos \left( {{q_x}} \right)\cos \left( {{q_y}} \right) + J_{01}^y{{\cos }^2}\left( {{q_y}} \right)} \right)
\label{eq:gap_coeff}
\end{eqnarray}
\end{widetext}
we finally get expressions for the finding of the superconducting gap components:
\begin{equation}
{\Delta _d} = \sum\limits_{\bf{q}} {{\Delta _{d{\bf{q}}}}}  = \sum\limits_{\bf{q}} {\left[ {{D_{1{\bf{q}}}}{\Delta _d} + {D_{2{\bf{q}}}}{\Delta _{s*}}} \right]},
\label{eq:gap_d}
\end{equation}
\begin{equation}
{\Delta _{s*}} = \sum\limits_{\bf{q}} {{\Delta _{s*{\bf{q}}}}}  = \sum\limits_{\bf{q}} {\left[ {{S_{1{\bf{q}}}}{\Delta _d} + {S_{2{\bf{q}}}}{\Delta _{s*}}} \right]}.
\label{eq:gap_s}
\end{equation}
The calculation of $d$- and $s^*$-wave components for each temperature and doping value is performed using a self-consistent solution of the system of Eqs.~(\ref{eq:gap_d}),(\ref{eq:gap_s}) together with the equation for the chemical potential and filling numbers.

\section{\label{sec:dsgaps}The concentration and temperature dependences of ${\bf{D}}$- and extended ${\bf{S^{\ast}}}$-wave components of the superconducting gap}

\begin{figure*}
\includegraphics[width=0.45\linewidth]{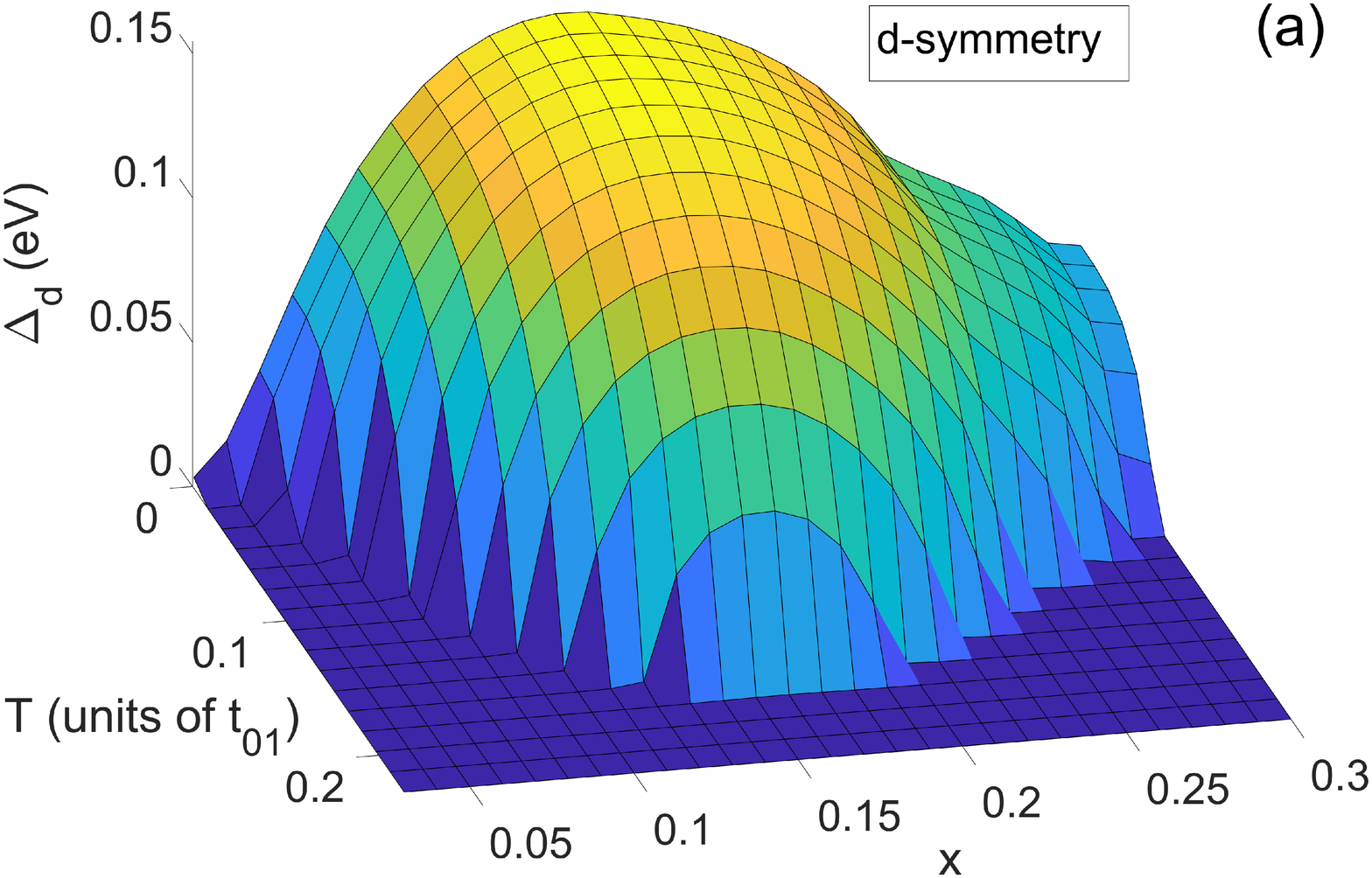}
\includegraphics[width=0.45\linewidth]{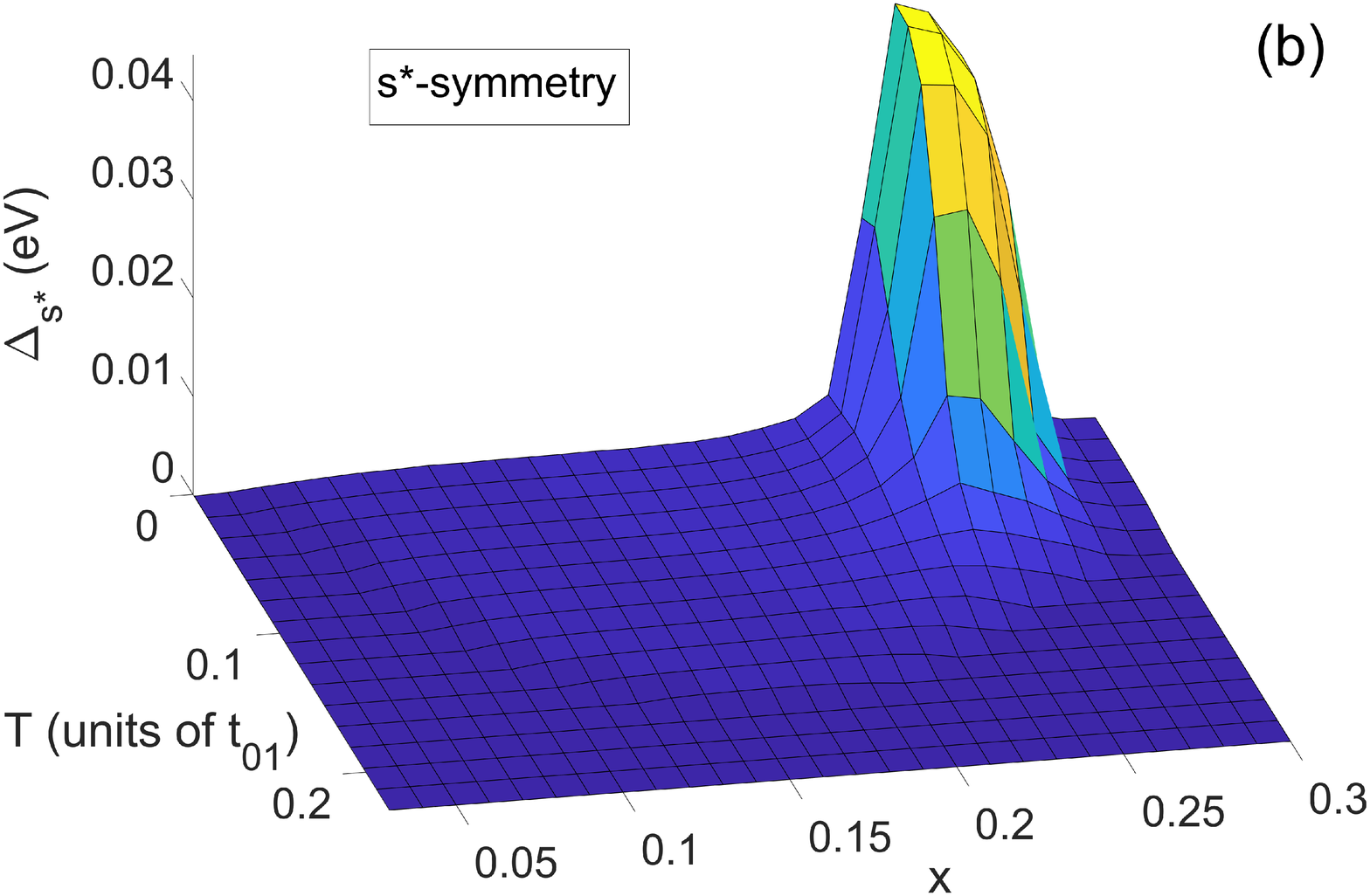}
\caption{\label{fig:d_s_gaps} The magnitudes of superconducting gap components of (a) $d$- and (b) extended $s^*$-wave symmetry in the space of parameters "doping-temperature" $\left( {x,T} \right)$ at orthorhombic distortion $\delta b /a = 4.15\% $. The temperature $T$ is in units ${t_{01}}$ (the interband quasiparticle hopping integral between the nearest sites of the crystal lattice along $x$ axis, $t_{01}=0.57$ eV).}
\end{figure*}

\begin{figure*}
\includegraphics[width=0.45\linewidth]{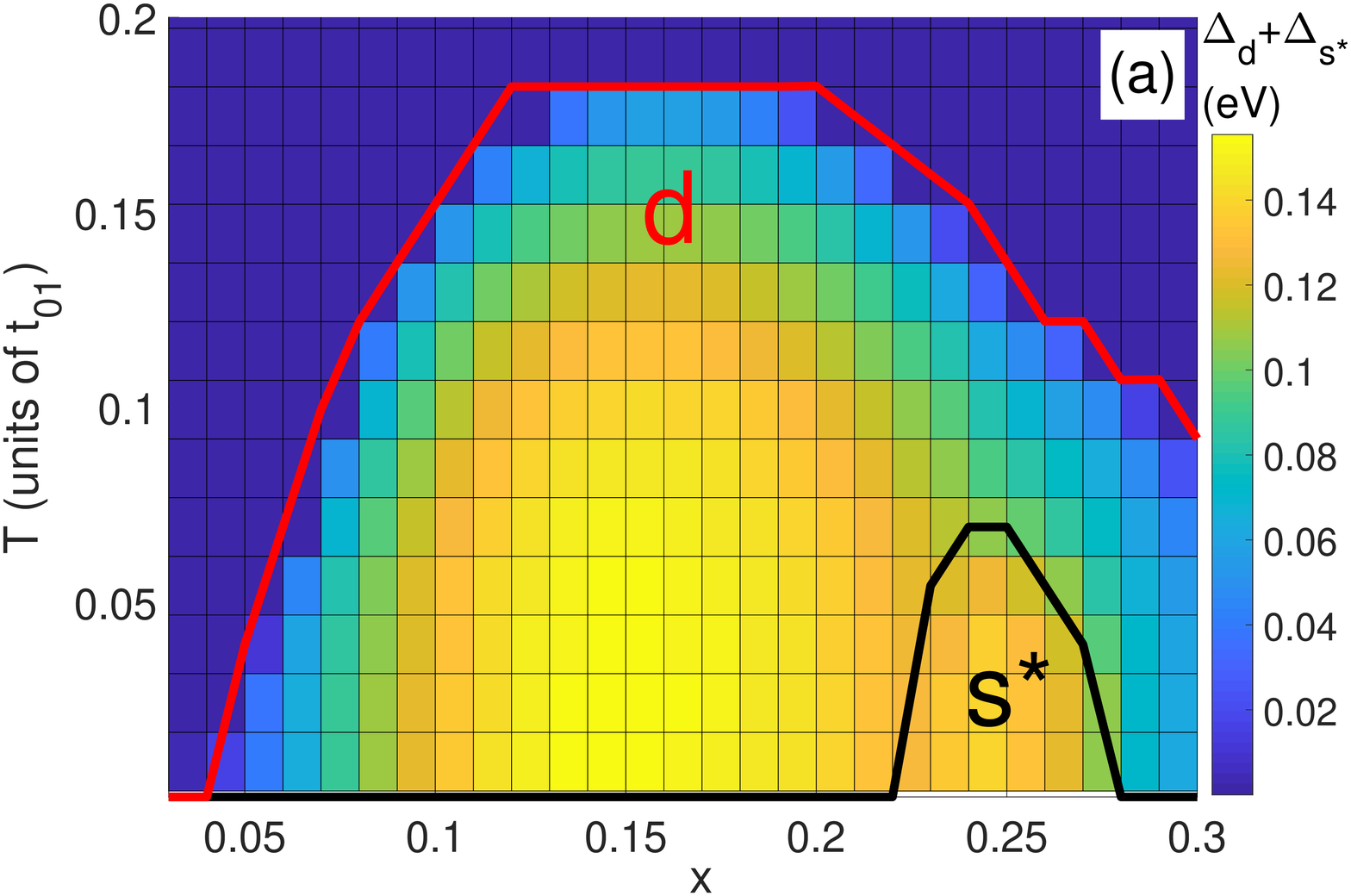}
\includegraphics[width=0.45\linewidth]{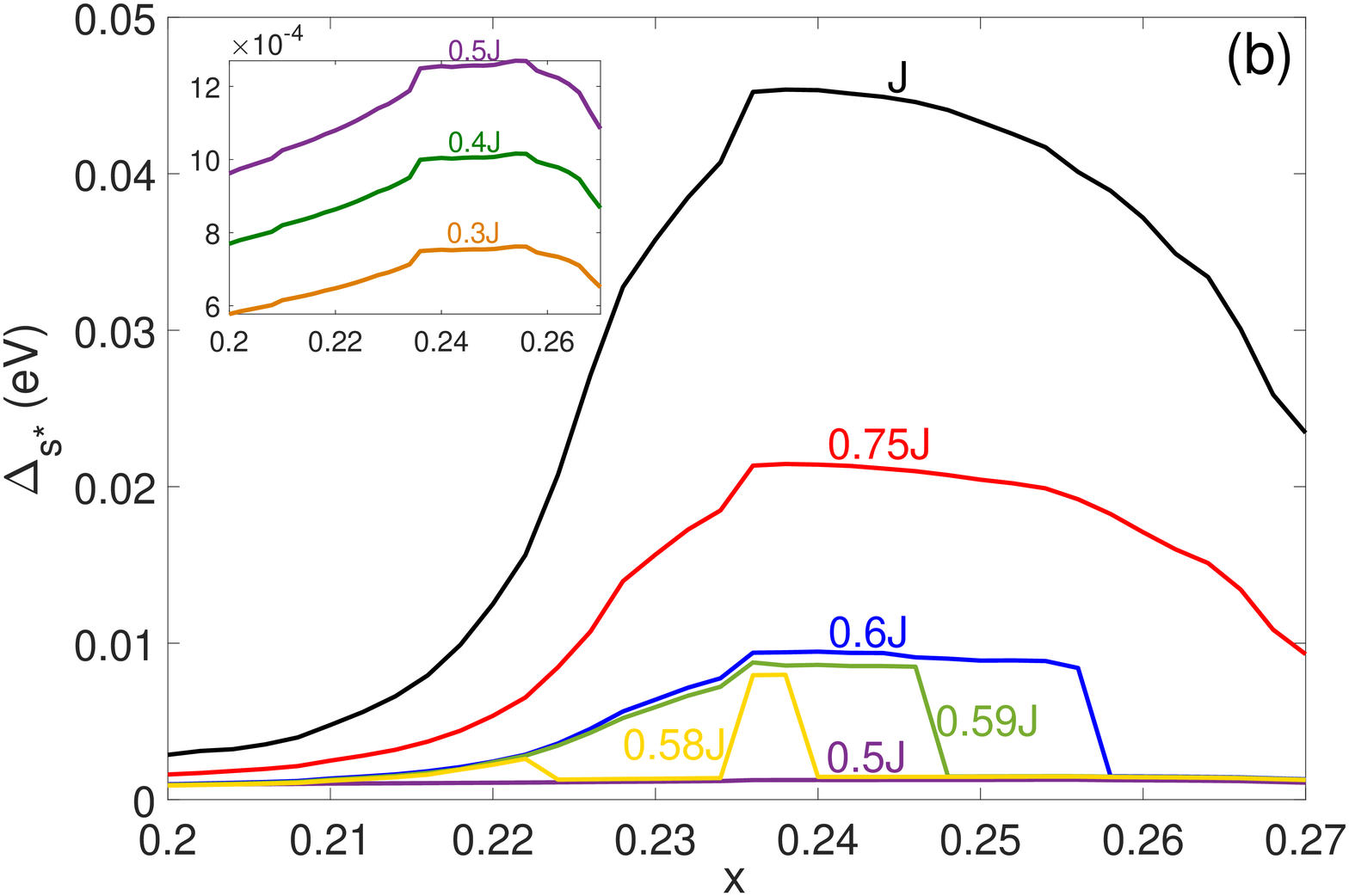}
\caption{\label{fig:phasediag} The diagram of $d$-wave and $s^*$-wave symmetry components of the superconducting gap in "doping-temperature" space of parameters in the system with orthorhombic distortion. The temperature $T$ is in units ${t_{01}}$ (the interband quasiparticle hopping integral between the nearest sites of the crystal lattice along $x$ axis, $t_{01}=0.57$ eV). (b) The region of the enhanced $s^*$-wave component of the superconducting gap upon varying the exchange pairing constant, the variable constant $J'$ varies from initial values $J$ (with $J_{01}^x$ and $J_{01}^y$) to $0.3J$. The inset in (b) demonstrates minor changes in the $s^*$-wave component at exchange interaction $J' \le 0.5J$.}
\end{figure*}
The temperature and concentration dependences of the $d$-wave symmetry and $s^*$-wave symmetry components are shown in Fig.~\ref{fig:d_s_gaps}a,b. The main contribution to the superconducting gap in cuprates is made by the $d$-wave component. The extended $s^*$-wave component is an additional component in the system with orthorhombic distortion. It is seen that the $d$-wave component prevails under the extended $s^*$-wave symmetry at all doping $x$ and temperature $T$ values. The dependence of the $d$-wave symmetry component on $x$ is different for various $T$. At $T < 0.1{t_{01}}$ (${t_{01}} \equiv {t_{\left( {01} \right)\sigma }}\left( {\sigma 0,\bar \sigma S} \right)$ is the interband quasiparticle hopping integral between the nearest sites of the crystal lattice along $x$ axis, $t_{01}=0.57$ eV), the behavior of ${\Delta _d}$ reminds the shape of the superconducting dome ${T_c}\left( x \right)$ obtained within the Hubbard model for the orthorhombic phase~\cite{Makarov21}. At $0.1{t_{01}} < T < 0.19{t_{01}}$, ${\Delta _d}\left( x \right)$ has a parabolic-like shape with the maximum between ${x_{c1}}$ and ${x_{c2}}$ (Fig.~\ref{fig:d_s_gaps}a), where ${x_{c1}}$ and ${x_{c2}}$ are concentrations of the first and second quantum phase transitions during which Fermi contour transforms from four hole pockets to the large hole and large electrons contours. The $T$ increasing results in ${\Delta _d}\left( x \right)$ decreasing, the gap closes at a critical temperature depending on doping. The contour of the surface ${\Delta _d}\left( {x,T} \right)$ at ${\Delta _d} = 0$ forms concentration dependence of ${T_c}$.

The $s^*$-wave component has a small value in the underdoped and optimally doped regions, its magnitude is defined by the small value of orthorhombic distortion. An interesting and unusual property is the sharp increase in the magnitude of the $s^*$-wave component in a narrow range of concentrations in the overdoped region from $x = 0.22$ to $x = 0.27$ (Fig.~\ref{fig:d_s_gaps}b) while the $d$-wave component simply decreases monotonically. The surface ${\Delta _{s*}}\left( {x,T} \right)$ has a sharp dome shape in this doping interval region (Fig.~\ref{fig:d_s_gaps}b). An increase in the $s^*$-wave component in comparison with the $d$-wave component results in a significant change in the reconstruction of the lines of the zeros of the superconducting gap. The gap nodes almost coincide with the nodal directions $\left( {0,0} \right) - \left( {2\pi ,2\pi } \right)$ and $\left( {2\pi ,0} \right) - \left( {0,2\pi } \right)$ at $x = 0.21$, they bend in such a way that the zero of the superconducting gap disappears at the ${\bf{k}}$-point $\left( {\pi ,\pi } \right)$ (Fig.~\ref{fig:Fermi_cont}a, red dotted lines). The lines of gap zeros are even more pushed apart in the region of the Brillouin zone center with a further increase in $x$ (Fig.~\ref{fig:Fermi_cont}b-d). The magnitude of the $s^*$-wave component decreases with temperature increasing (Fig.~\ref{fig:d_s_gaps}b). Reduction of the $s^*$-wave component with temperature increasing down to 0 is more gradual compared to the decrease in ${\Delta _d}$.  The $s^*$-wave component magnitude reaches half of the $d$-wave component in its maximum at $x = 0.234$ and $T = 0$ K. It is obvious that a new mechanism of the influence of orthorhombic distortion on the superconducting gap begins to work at strong doping. Location of the enlarged $s^*$-wave component region on the phase diagram $\left( {x,T} \right)$ is depicted in Fig.~\ref{fig:phasediag}. The narrow range of doped hole concentrations at which the calculated $s^*$-wave component is enlarged is located just behind the concentration of the pseudogap closing ${p^*} = 0.22$ in the common phase diagram based on experimental data~\cite{Hashimoto14}. To elucidate the reason for the sharp growth of the $s^*$-wave component we will study in Sections~\ref{sec:elstr} and~\ref{sec:partcontr} in more detail how the transformation of the electronic structure in the doping range from $x = 0.21$ to $x = 0.27$ can change the amount and the size of contributions of various electronic states to superconducting gap with different symmetry.

The exchange interaction constants $J_{01}^x$ and $J_{01}^y$ which determine the value of the pairing interaction in the approach used are close to the values obtained from experiments on inelastic neutron scattering in cuprates~\cite{Coldea01} and from calculations that take into account the complete basis of quasiparticle cluster excitations~\cite{Sidorov16} ($J = 0.149$ eV). To check whether the effect of the $s^*$-wave component increasing remains at lower exchange interaction constants its dependence on doping in the $x$ range from $0.21$ to $0.27$ was calculated with artificially reduced $J$ (Fig.~\ref{fig:phasediag}b), the reduced constant $J$ is denoted as $J'$. The hump of the $s^*$-wave component gradually subsides when $J$ decreasing to $60\%$ of the initial value. The value ${\Delta _{s*}}$ decreases very rapidly at $J' > 0.6J$ ($0.1$ eV). The effect of an increase in the $s^*$-wave component completely disappears at $J' = 0.5J$ ($0.08$ eV).
 
\section{\label{sec:elstr}Electronic structure of low-energy excitations and its transformation with doping in the overdoped region}

The lower and upper Hubbard electron subbands of cuprates in the normal phase are depicted in Fig.~\ref{fig:band_str}a,b (color lines). Two more hole branches (Fig.~\ref{fig:band_str}a,b, gray lines) are added when describing the superconducting phase using the Hubbard model in terms of the Gorkov-Nambu operators in quantum field theory. The bands of Bogolyubov quasiparticles are formed as a result of the mixing of the states of the electron and hole branches due to pairing. The gap equal to the doubled value of the superconducting gap opens between Bogolyubov quasiparticles bands. To understand the features of the concentration dependence of the gap, it is necessary to clarify what characteristic changes occur in the electronic structure of quasiparticles in the normal phase and Bogolyubov quasiparticles in the superconducting phase with doping. The electronic structure of the cuprate with orthorhombic distortion in the normal phase in the wide doping range was obtained in work~\cite{Makarov21}. Here we will analyze in more detail the electronic structure in the normal and superconducting phases in the narrow doping range from $x = 0.21$ to $x = 0.24$ in which significant changes in the ratio of the components of the superconducting gap occur.

\subsection{\label{sec:elstr_ns}Transformation of the electronic structure with doping in the normal state}

\begin{figure*}
\includegraphics[width=0.45\linewidth]{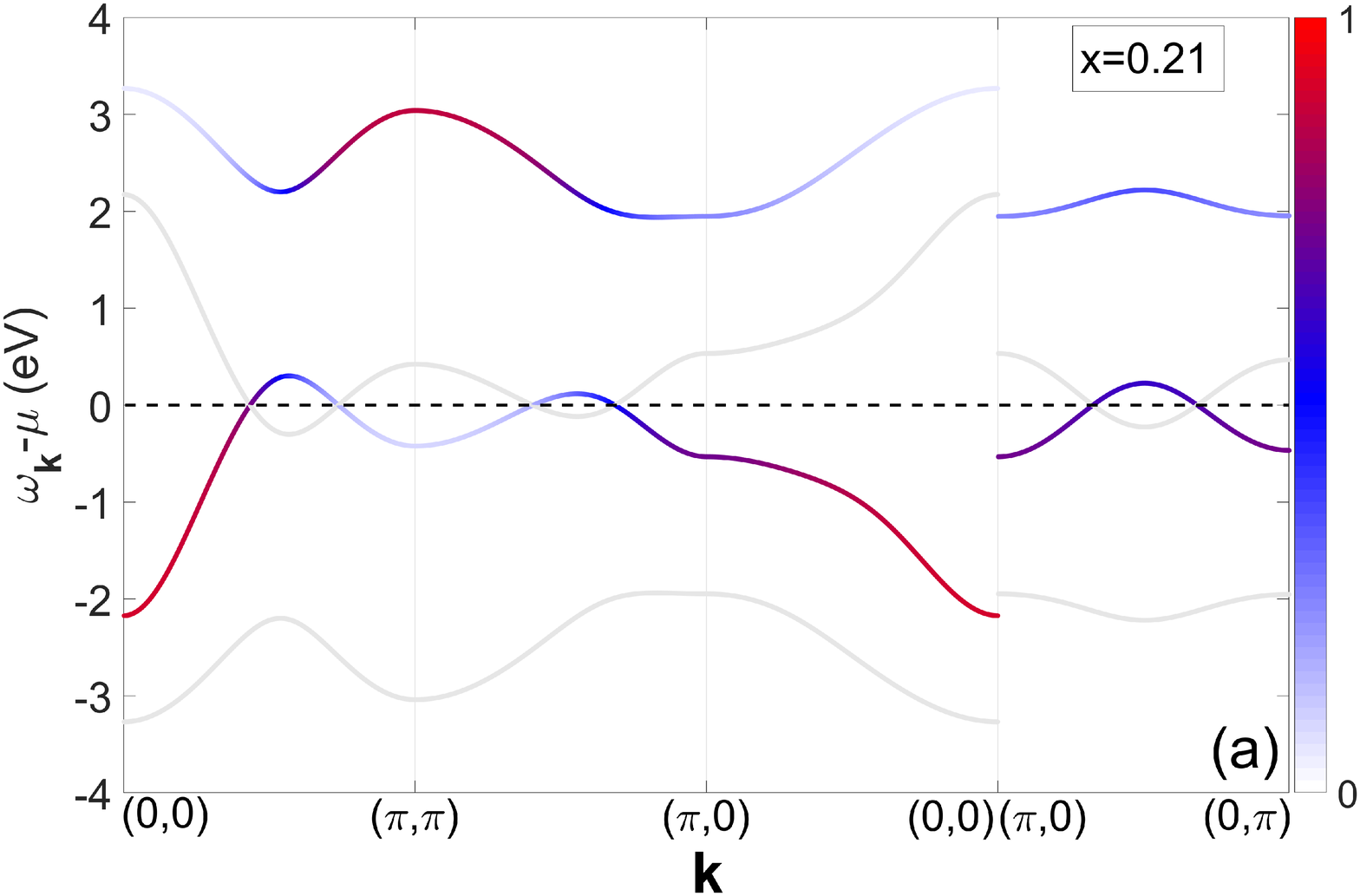}
\includegraphics[width=0.45\linewidth]{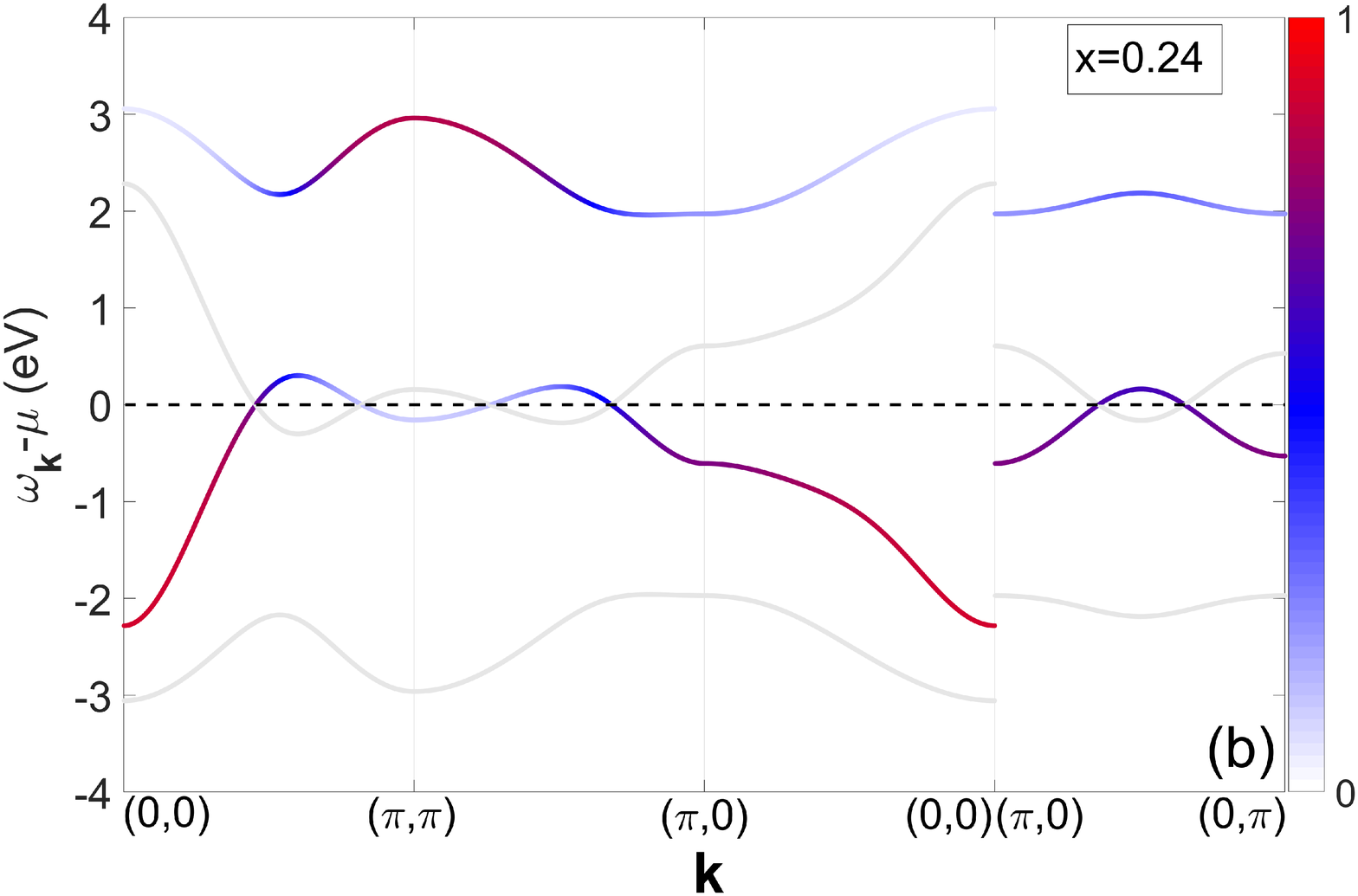}
\includegraphics[width=0.45\linewidth]{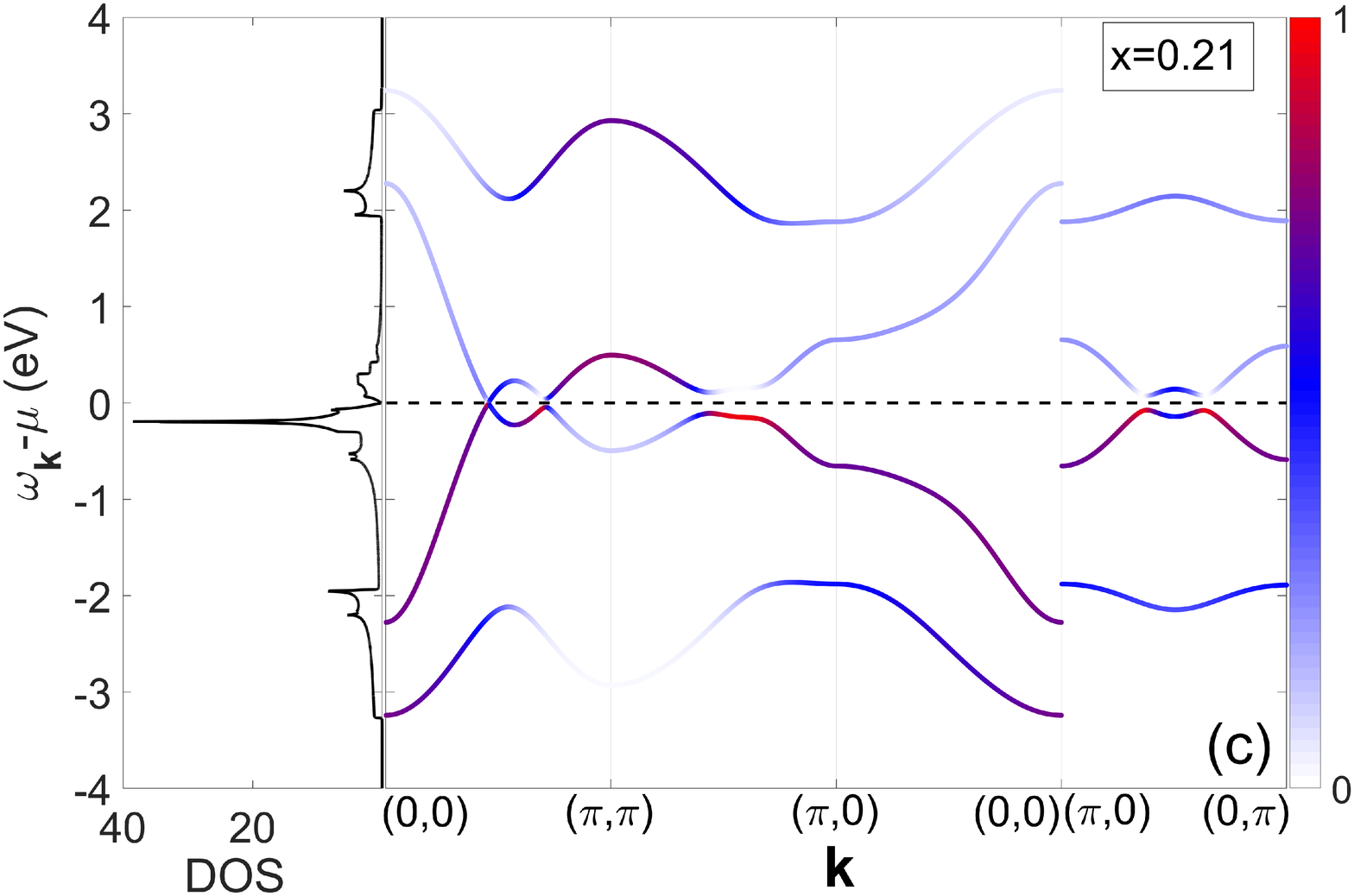}
\includegraphics[width=0.45\linewidth]{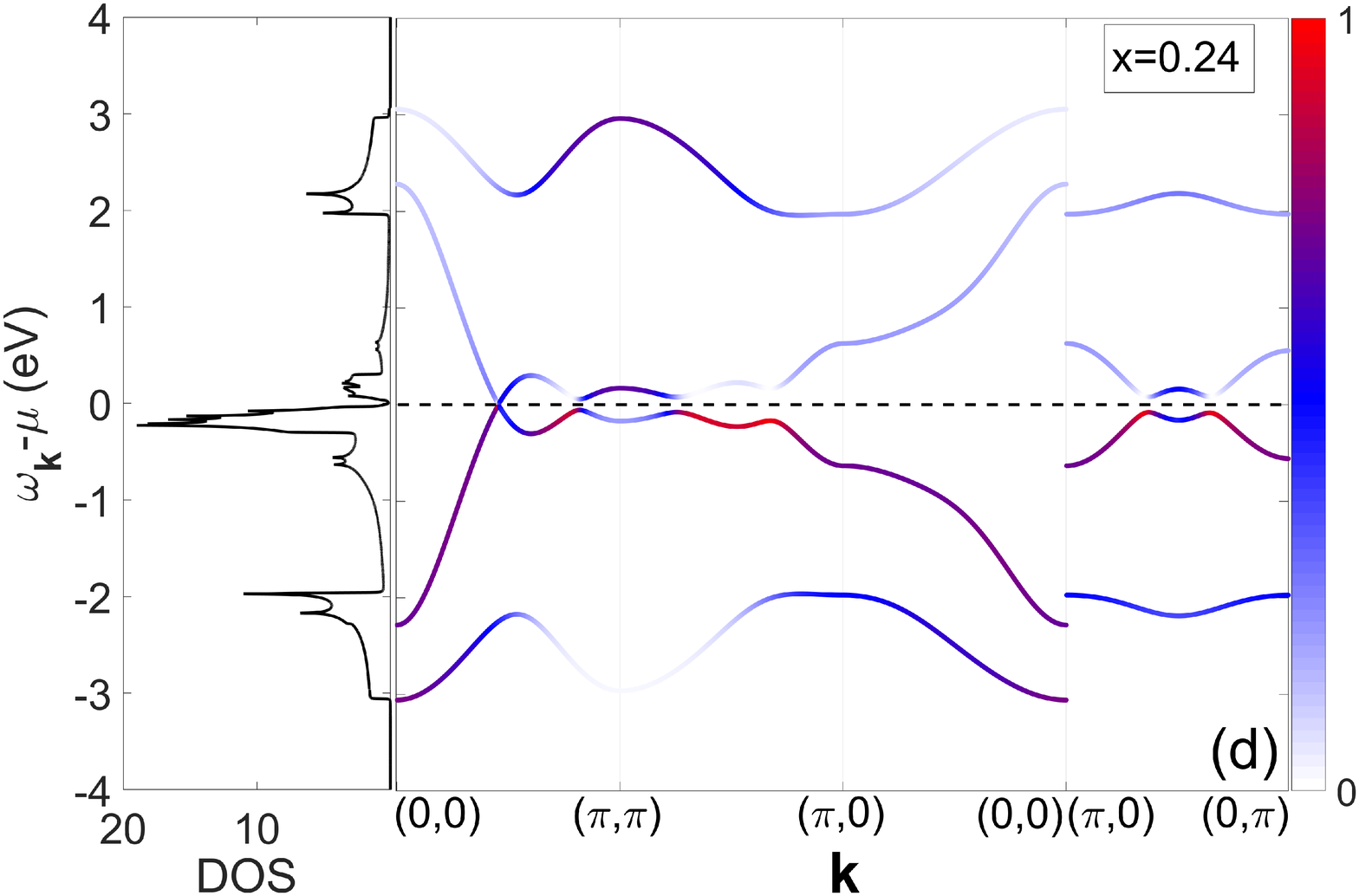}
\caption{\label{fig:band_str} The evolution of the band structure of the quasiparticles in the normal phase (a,b) and Bogolyubov quasiparticles in the superconducting phase (c,d) calculated within the Hubbard model for orthorhombic cuprate with doping from $x = 0.21$ to $0.24$. The color of each dispersion point indicates the spectral weight of the state with the corresponding wave vector ${\bf{k}}$. The gray line in (a,b) shows the hole branch of the Bogolyubov quasiparticle band in the system without superconducting pairing. The left panels of (c,d) show the density of states of Bogolyubov quasiparticles in the superconducting phase.}
\end{figure*}

\begin{figure*}
\includegraphics[width=0.6\linewidth]{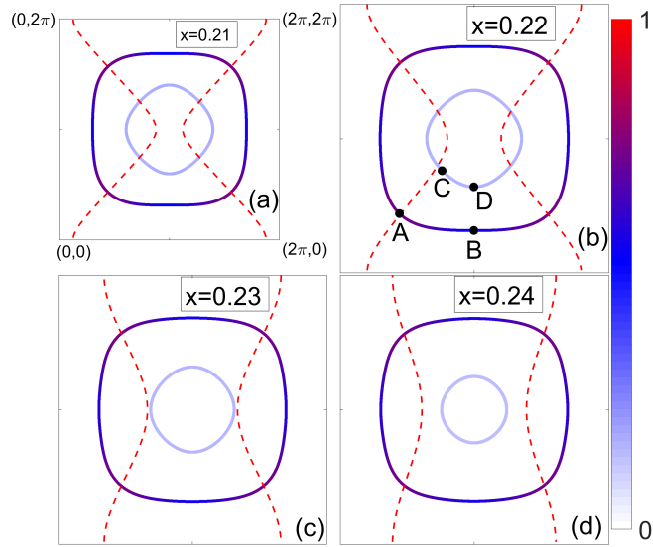}
\caption{\label{fig:Fermi_cont} The evolution of the Fermi contours and the superconducting gap nodes with doping $x$. The Fermi contour is represented by two contours around the point $\left( {\pi ,\pi } \right)$, the large outer contour consists of the states with a higher spectral weight, the smaller inner electron contour consists of the states with a lower intensity. The dotted red lines are the superconducting gap nodes. It can be seen that the nodes go beyond the small electron contour and stops crossing it with an increase in the hole concentration between $x = 0.22$ and $x = 0.23$.}
\end{figure*}

First, we will consider the electronic structure in the normal phase, namely the structure of the lower Hubbard zone (LHB), since the chemical potential is located inside this band, and the states of this band will determine the electronic properties. LHB in the orthorhombic phase is asymmetric with respect to reflections in planes passing through the $\left( {0,0} \right) - \left( {{{2\pi } \mathord{\left/
 {\vphantom {{2\pi } a}} \right.
 \kern-\nulldelimiterspace} a},{{2\pi } \mathord{\left/
 {\vphantom {{2\pi } b}} \right.
 \kern-\nulldelimiterspace} b}} \right)$ and $\left( {{{2\pi } \mathord{\left/
 {\vphantom {{2\pi } a}} \right.
 \kern-\nulldelimiterspace} a},0} \right) - \left( {0,{{2\pi } \mathord{\left/
 {\vphantom {{2\pi } b}} \right.
 \kern-\nulldelimiterspace} b}} \right)$ directions perpendicular to the ${k_x} - {k_y}$ plane~\cite{Makarov21}. Further in the article, we will omit the multiplier ${1 \mathord{\left/
 {\vphantom {1 a}} \right.
 \kern-\nulldelimiterspace} a}$ when specifying the ${k_x}$ component of the wave vector and the multiplier ${1 \mathord{\left/
 {\vphantom {1 b}} \right.
 \kern-\nulldelimiterspace} b}$ when specifying its ${k_y}$ component in order not to overload the text with a large number of symbols. LHB has absolute maxima near points  $\left( {{\pi  \mathord{\left/
 {\vphantom {\pi  2}} \right.
 \kern-\nulldelimiterspace} 2},{\pi  \mathord{\left/
 {\vphantom {\pi  2}} \right.
 \kern-\nulldelimiterspace} 2}} \right)$, $\left( {{{3\pi } \mathord{\left/
 {\vphantom {{3\pi } 2}} \right.
 \kern-\nulldelimiterspace} 2},{\pi  \mathord{\left/
 {\vphantom {\pi  2}} \right.
 \kern-\nulldelimiterspace} 2}} \right)$, $\left( {{\pi  \mathord{\left/
 {\vphantom {\pi  2}} \right.
 \kern-\nulldelimiterspace} 2},3{\pi  \mathord{\left/
 {\vphantom {\pi  2}} \right.
 \kern-\nulldelimiterspace} 2}} \right)$, $\left( {{{3\pi } \mathord{\left/
 {\vphantom {{3\pi } 2}} \right.
 \kern-\nulldelimiterspace} 2},{{3\pi } \mathord{\left/
 {\vphantom {{3\pi } 2}} \right.
 \kern-\nulldelimiterspace} 2}} \right)$, local maxima at points of directions $\left( {\pi ,0} \right) - \left( {\pi ,2\pi } \right)$, $\left( {0,\pi } \right) - \left( {2\pi ,\pi } \right)$, and a local minimum at point $\left( {\pi ,\pi } \right)$ (Fig.~\ref{fig:band_str}a,b). The distribution of spectral weight over states with different ${\bf{k}}$ is inhomogeneous, the value of the spectral weight is shown by color in Fig.~\ref{fig:band_str}a,b. The chemical potential in the doping range from $x = 0.21$ to $x = 0.24$ lies between the local maxima and the local minimum. As a result, the large outer Fermi contour and the small inner electron Fermi contour are formed around the point $\left( {\pi ,\pi } \right)$ (Fig.~\ref{fig:Fermi_cont})~\cite{Makarov21}. This is precisely the topology of the Fermi contour in the doping range of interest to us. The states on the outer contour have higher spectral intensity in comparison with the states on the inner electron contour.
 The most significant changes in the band structure with an increase in the concentration of holes in the region of strong doping are (i) a decrease in the depth of the inner electron pocket and (ii) contraction of this pocket. The effect of the reduction of the small inner electron pocket depth is visible when comparing the band structures in Fig.~\ref{fig:band_str}a,b. It is seen that the deepening of the dispersion surface around the point $\left( {\pi ,\pi } \right)$ becomes smaller as the doping changes from $x = 0.21$ to $x = 0.24$. This effect is a consequence of the weakening of spin correlations. An increase in the number of holes results in a downward shift of the chemical potential to the valence band. Doping-induced reconstruction of the dispersion and the shift of the chemical potential lead to a decrease in the outer and inner Fermi contours and a change in their shape (Fig.~\ref{fig:Fermi_cont}a-d). The shape of the outer contour changes insignificantly, as does its size; contraction of this pocket is maximal along with the nodal directions $\left( {0,0} \right) - \left( {2\pi ,2\pi } \right)$ and $\left( {0,2\pi } \right) - \left( {2\pi ,0} \right)$, and is almost absent along the antinodal directions $\left( {\pi ,0} \right) - \left( {\pi ,2\pi } \right)$ and $\left( {0,\pi } \right) - \left( {2\pi ,\pi } \right)$ (Fig.~\ref{fig:Fermi_cont}a-d). The shape of the inner contour changes from rectangular with smooth corners at $x = 0.21$ (Fig.~\ref{fig:Fermi_cont}a) to a circle around states near ${\bf{k}} = \left( {\pi ,\pi } \right)$ at $x = 0.24$ (Fig.~\ref{fig:Fermi_cont}d). The rate of reduction of the doped inner Fermi contour with hole doping is greater than that for the outer Fermi contour. The number of states in the inner pocket decreases with hole doping from $12\%$ of the total Brillouin zone at $x = 0.21$ to $0\%$ at $x = 0.28$ (the inner contour disappears and only the outer contour remains).
\subsection{\label{sec:elstr_ss}Transformation of the electronic structure of Bogolyubov quasiparticles with doping in the superconducting state}
\begin{figure*}
\includegraphics[width=0.45\linewidth]{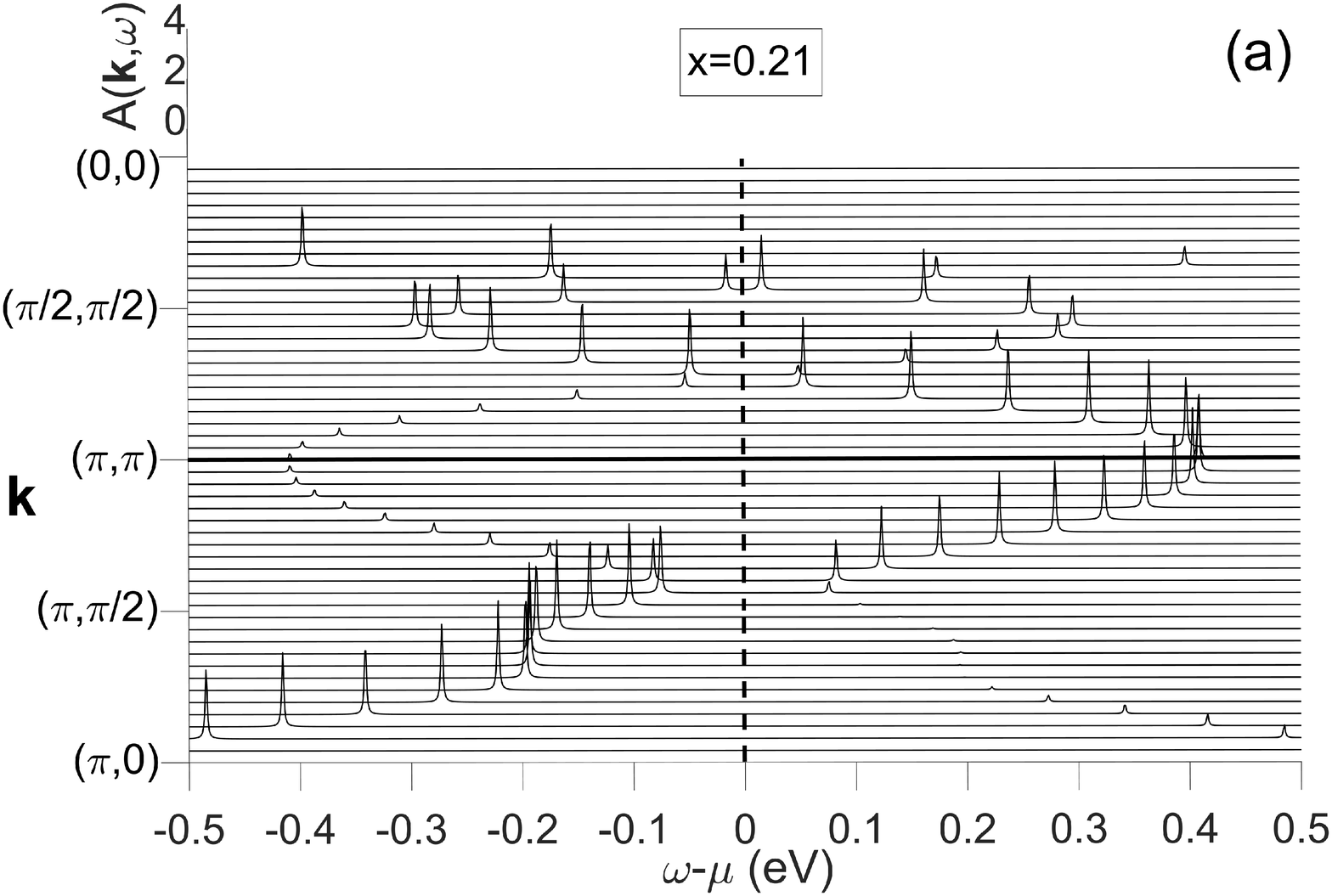}
\includegraphics[width=0.45\linewidth]{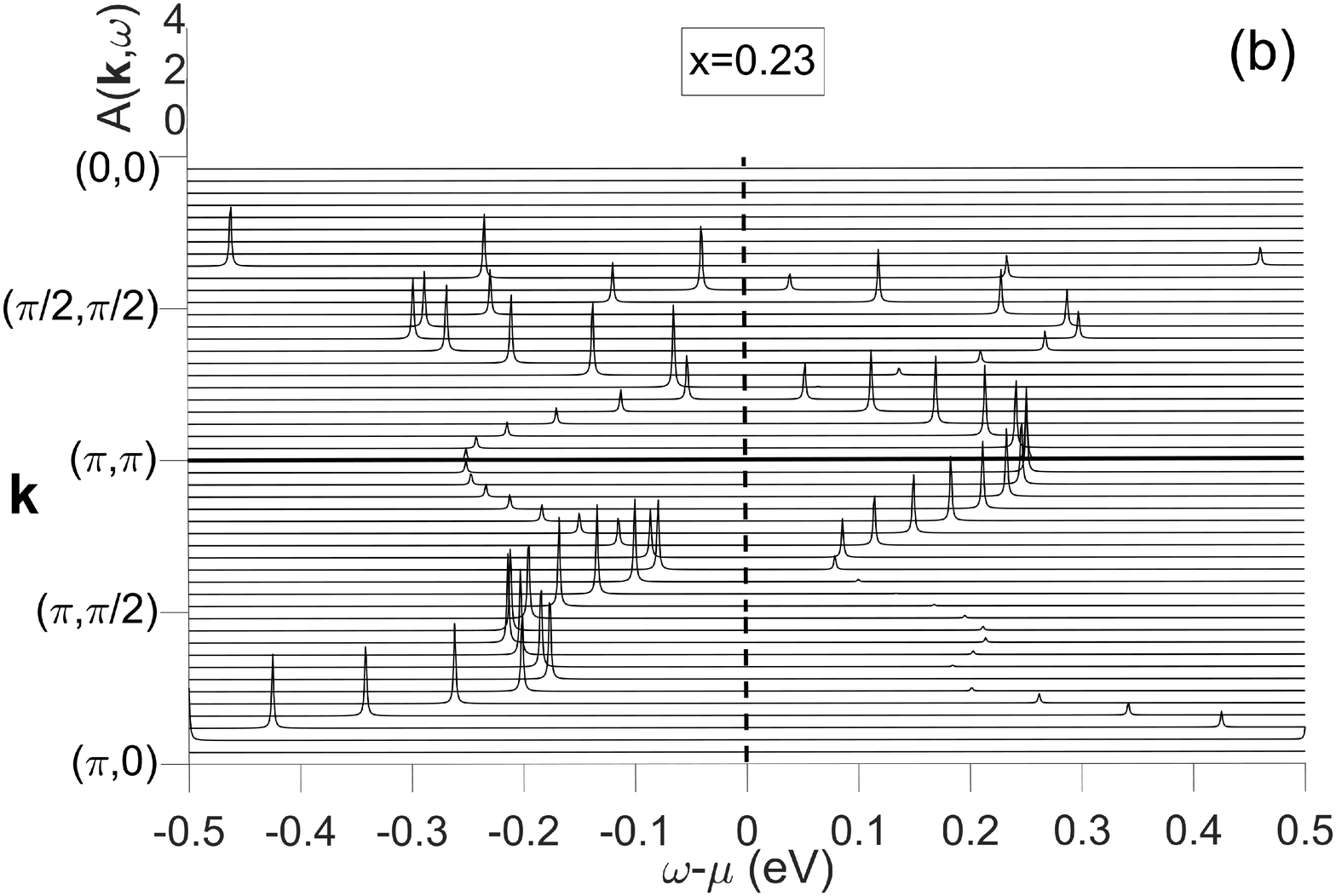}
\includegraphics[width=0.45\linewidth]{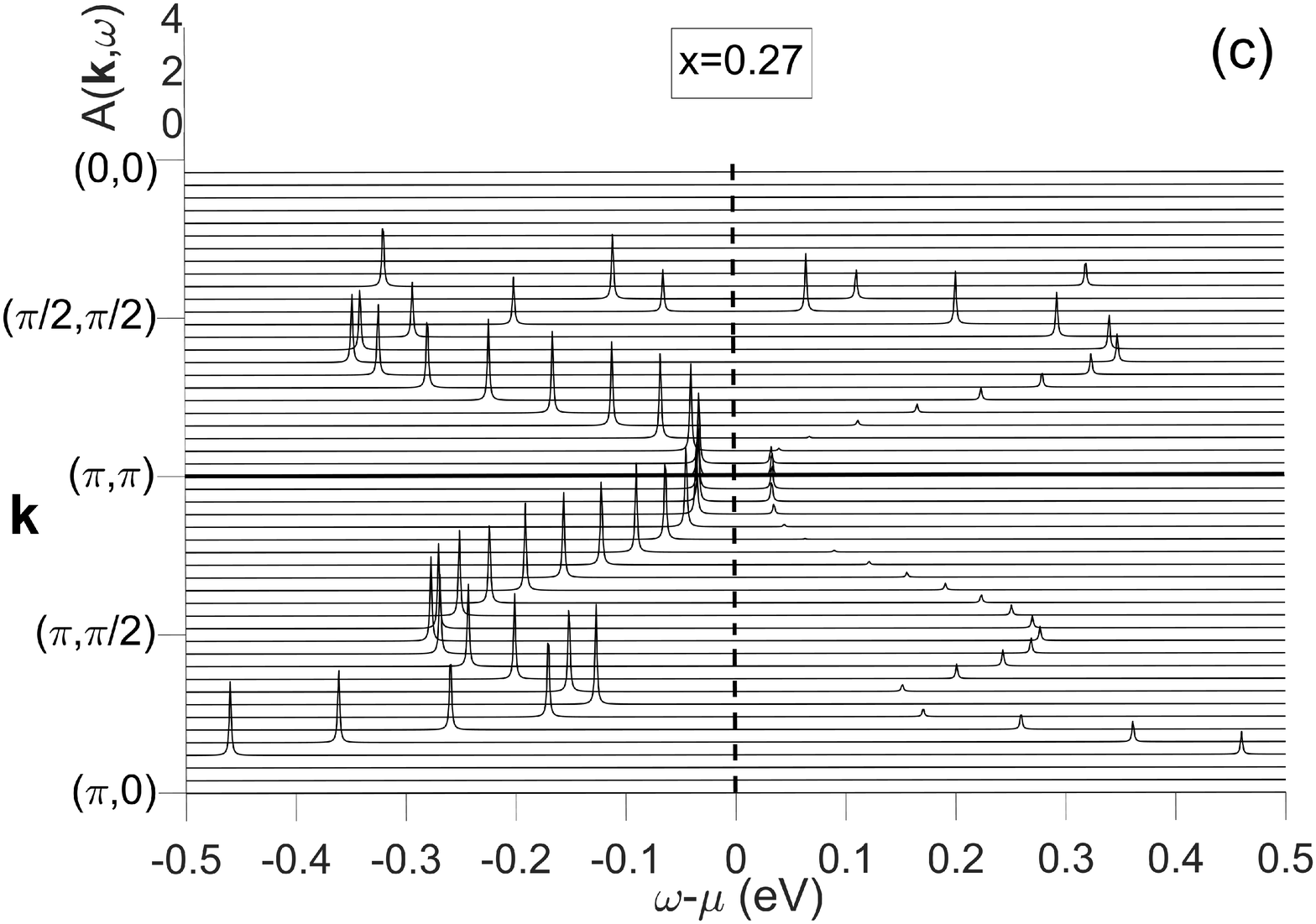}
\includegraphics[width=0.45\linewidth]{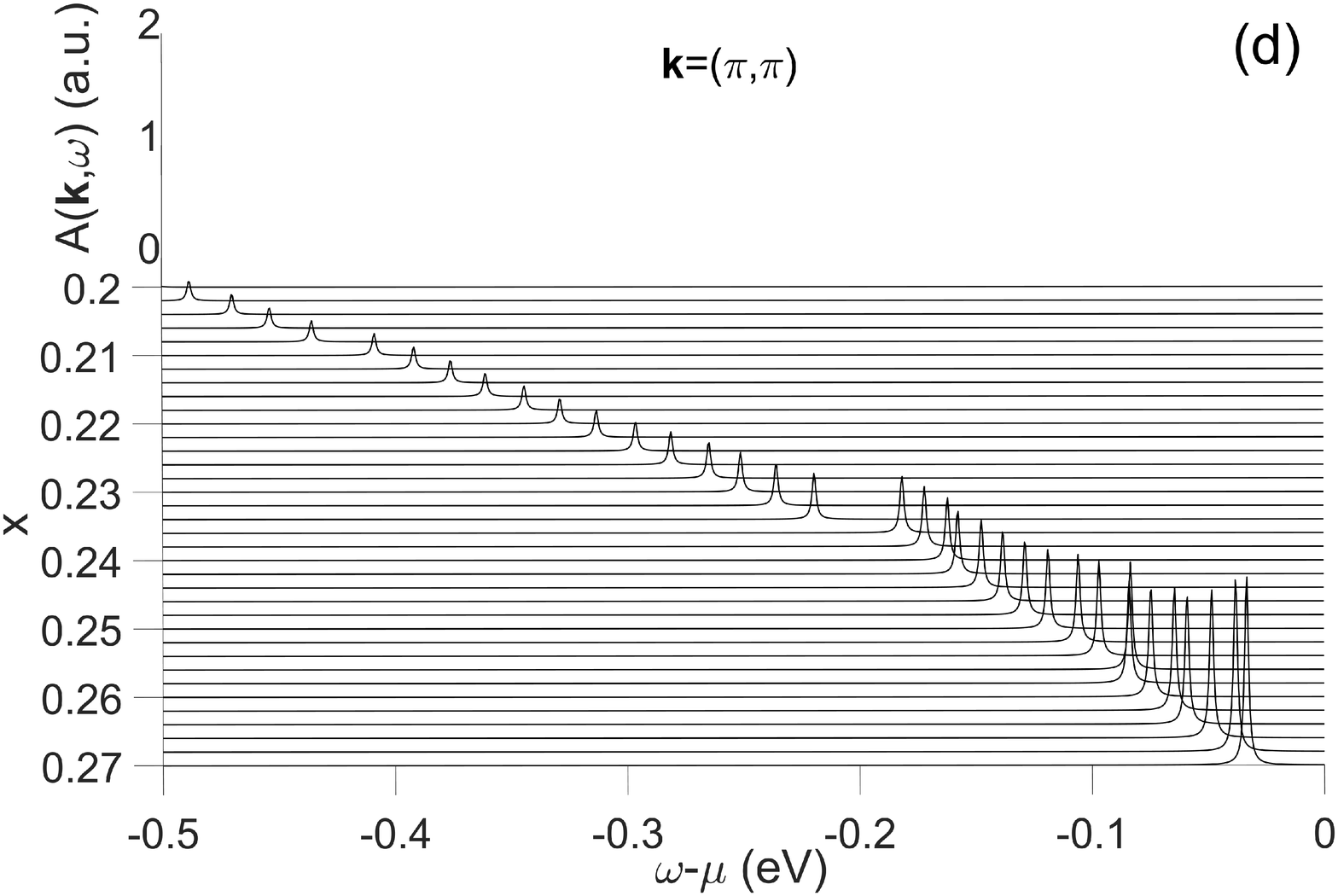}
\caption{\label{fig:lowenergy} The spectral functions of Bogolyubov quasiparticles in the region of low-energy excitations along the directions $\left( {0,0} \right) - \left( {\pi ,\pi } \right)$ and $\left( {\pi ,\pi } \right) - \left( {\pi ,0} \right)$ and their concentration dependence. The band structures are plotted without Fermi cutoff in contrast to the ARPES spectra which do not show empty states. (d) The concentration dependence of the spectral functions at ${\bf{k}} = \left( {\pi ,\pi } \right)$ which indicates a decrease in the depth of the inner pocket and growth of the spectral weight of states in it as $x$ increasing.}
\end{figure*}
\begin{figure*}
\includegraphics[width=0.45\linewidth]{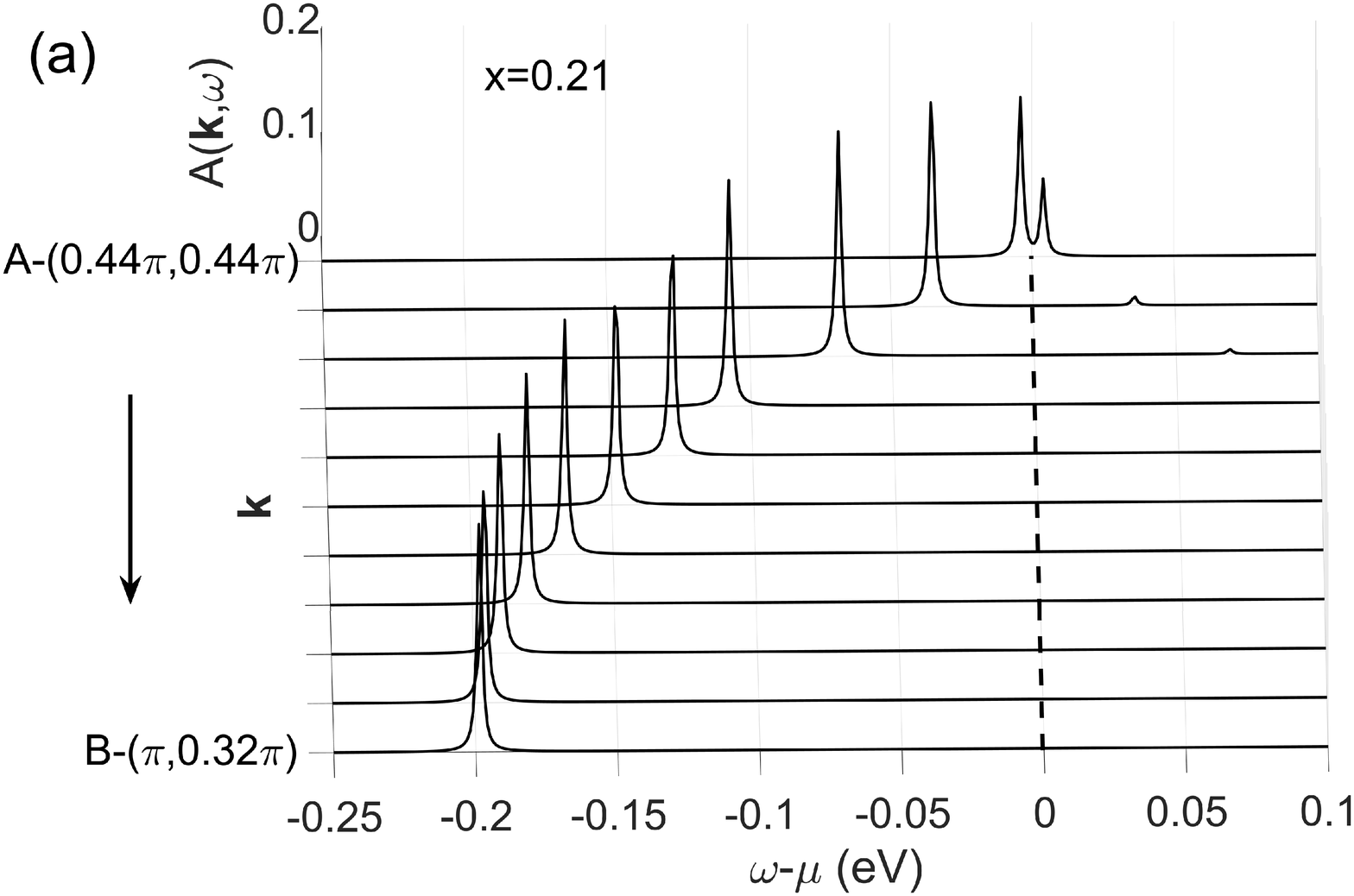}
\includegraphics[width=0.45\linewidth]{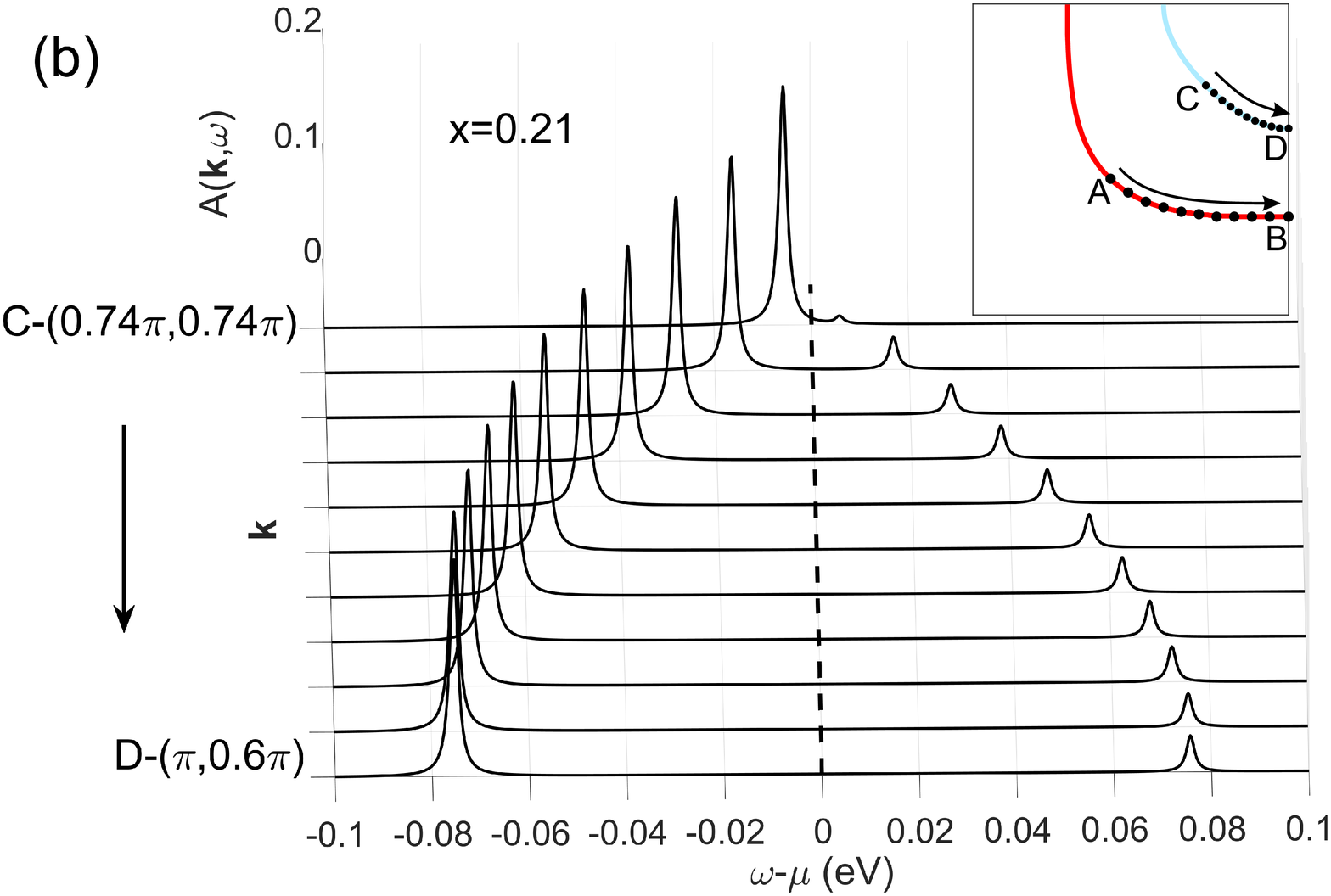}
\includegraphics[width=0.45\linewidth]{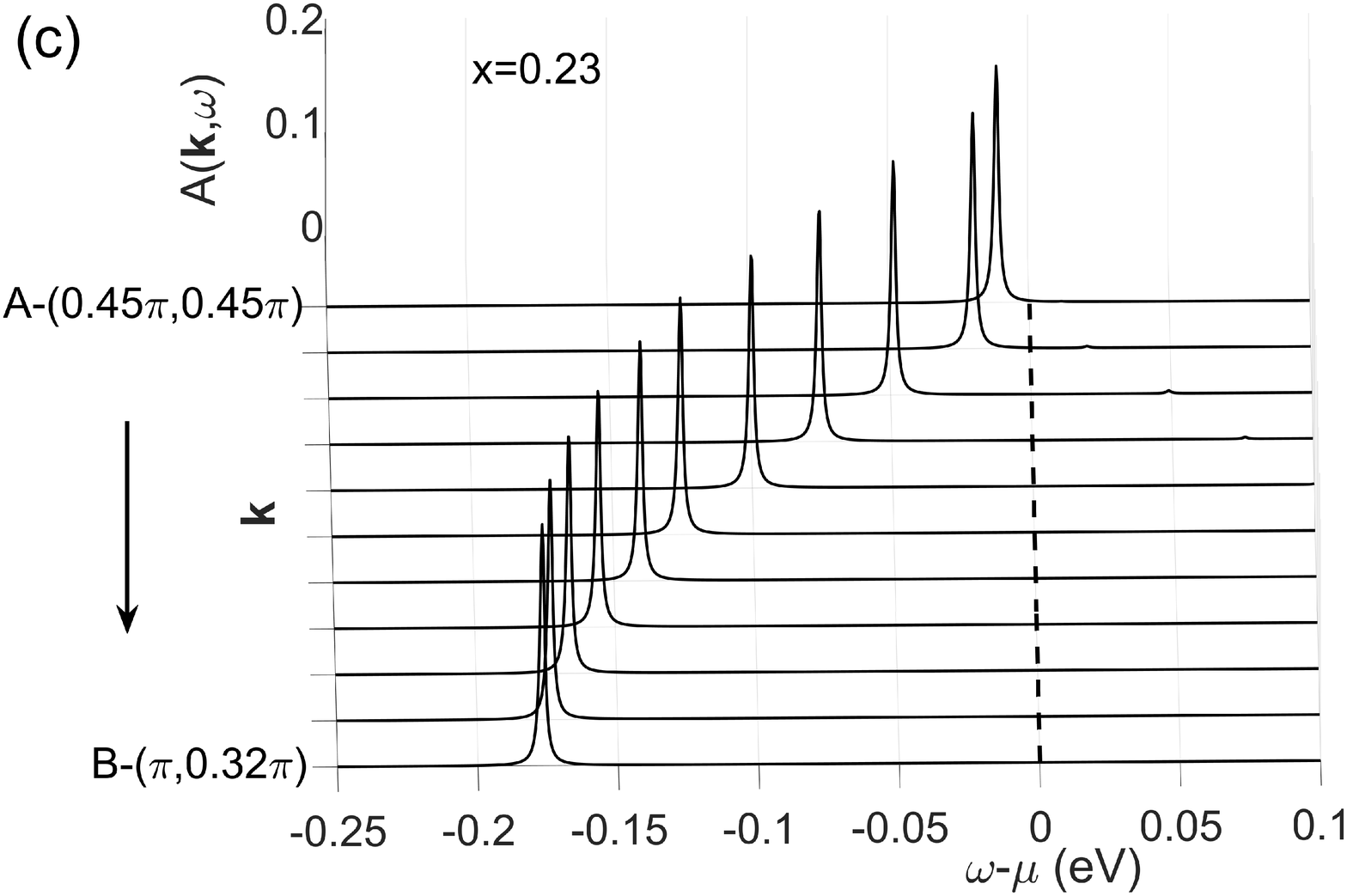}
\includegraphics[width=0.45\linewidth]{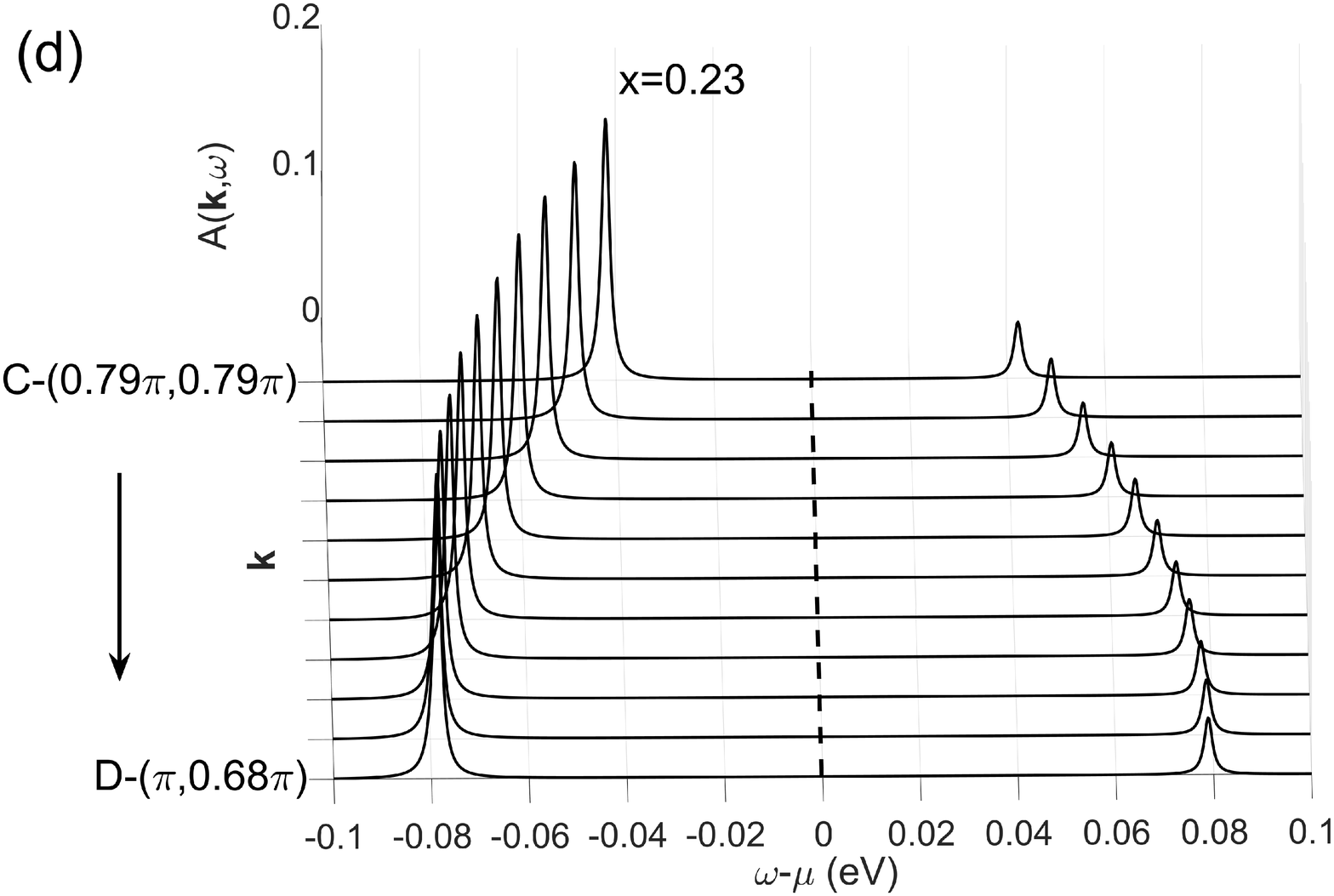}
\caption{\label{fig:spect_func} The spectral functions of Bogolyubov quasiparticles at ${{\bf{k}}_F}$ along large outer Fermi contour from point A to B (red line in the inset of (b)) and along small inner Fermi contour from point C to D (blue line in the inset of (b)) and their evolution with doping from $x=0.21$ to $x=0.23$.}
\end{figure*}
\begin{figure*}
\includegraphics[width=0.45\linewidth]{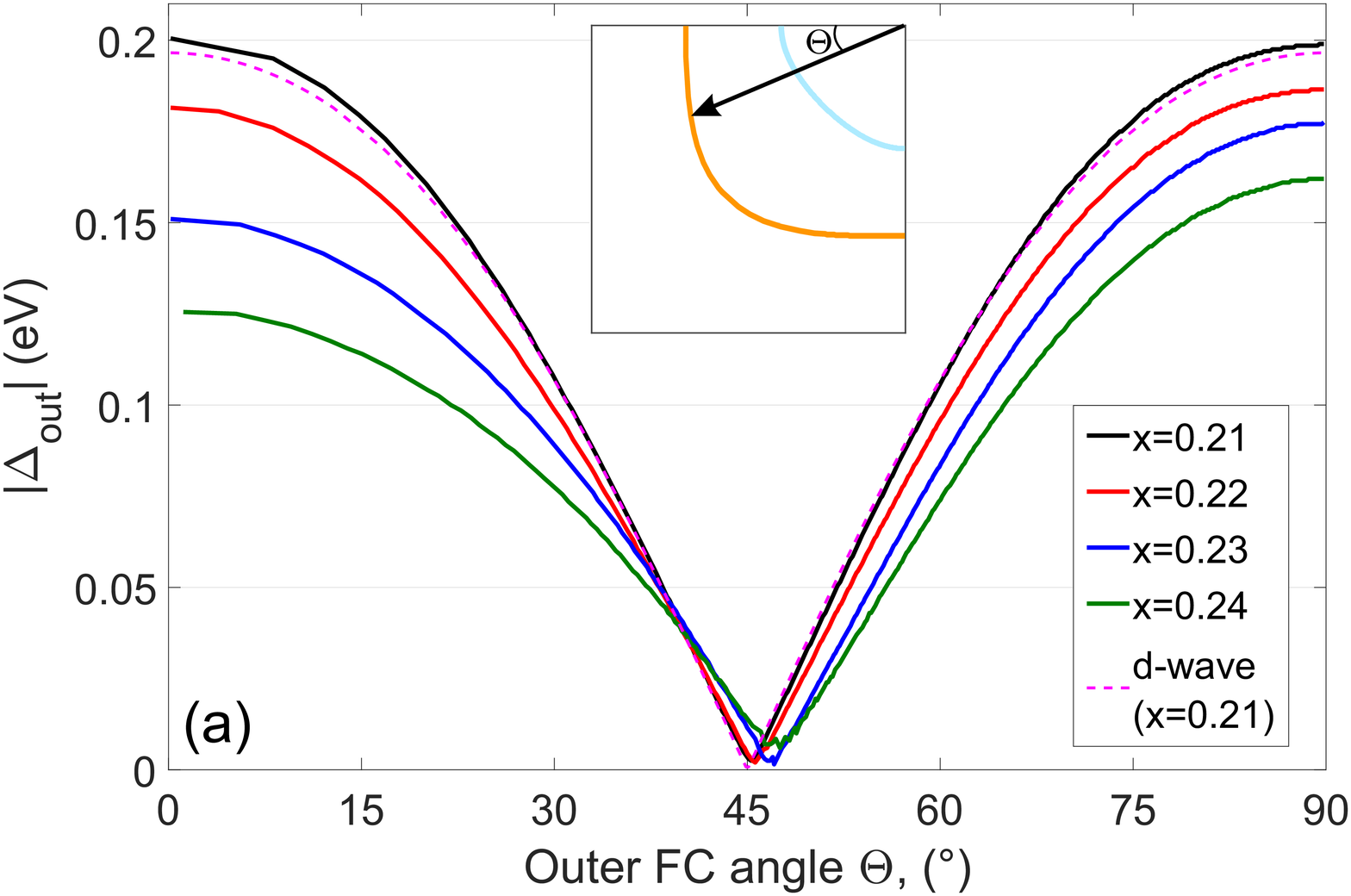}
\includegraphics[width=0.45\linewidth]{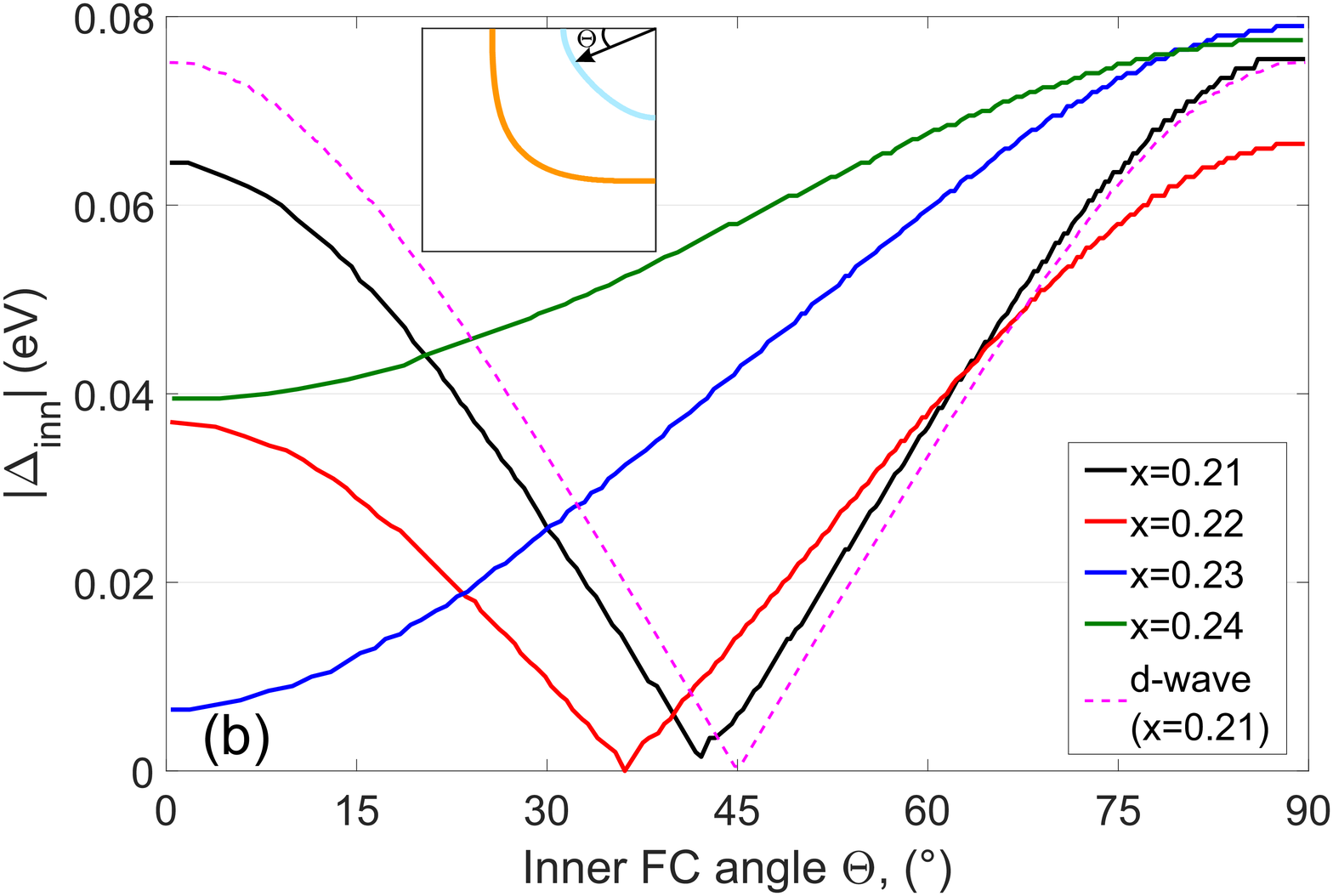}
\caption{\label{fig:gap_profile} The superconducting gap at ${{\bf{k}}_F}$ along (a) outer Fermi contour (orange line in the inset) and (b) inner Fermi contour (blue line in the inset) as a function of the Fermi angle $\Theta $ at different doping. The profiles of the superconducting gap are plotted from the position of the spectral function peaks along outer and inner contours at this hole concentration. The profile of gap along inner contour significantly changes between $x=0.22$ and $x=0.23$. The magenta dashed line shows the $d$-wave gap in the tetragonal phase at $x=0.21$.}
\end{figure*}
Now we consider the electronic structure in the superconducting phase and its dependence on doping. The total dispersion of Bogolyubov quasiparticles is shown in Fig.~\ref{fig:band_str}c,d. The spectral weight distribution of Bogolyubov quasiparticles is inhomogeneous. The upper Hubbard band and its hole branch will make a small contribution to the superconducting gap; therefore, it is worth analyzing two bands formed as a result of the hybridization of the lower Hubbard band and its hole branch. These bands are located in the energy range from $-2$ eV to $2$ eV. The dispersion in the interval of energies below $-0.5$ eV and above $0.5$ eV almost repeats the topology of the lower Hubbard band and its hole branch in the normal state. The bands of Bogolyubov quasiparticles within the energy of $0.5$ eV above and below the chemical potential $\mu$ are significantly renormalized due to pairing. The spectral functions of Bogolyubov quasiparticles in the region of low-energy excitations along the directions $\left( {0,0} \right) - \left( {\pi ,\pi } \right)$ and $\left( {\pi ,\pi } \right) - \left( {\pi ,0} \right)$, as well as their evolution with doping, are shown in Fig.~\ref{fig:lowenergy}. A decrease in the depth of the inner electron pocket around the ${\bf{k}}$-point $\left( {\pi ,\pi } \right)$ with doping is still the most significant variation of the electronic structure in the superconducting state. It can be seen that the peaks of the spectral functions of states inside the inner electron pocket near the ${\bf{k}}$-point $\left( {\pi ,\pi } \right)$ are shifted closer to the chemical potential when hole concentration grows from $x=0.21$ to $x=0.27$ (Fig.~\ref{fig:lowenergy}a-c). This effect is more clearly visible by the example of the shift of the peak of the spectral function at the wave vector $\left( {\pi ,\pi } \right)$ from $-0.4$ eV to $-0.05$ eV (Fig.~\ref{fig:lowenergy}d). An increase in the spectral weight of all states inside the inner pocket and, in particular, at the ${\bf{k}}$-point $\left( {\pi ,\pi } \right)$ with increasing $x$ is also clearly visible (Fig.~\ref{fig:lowenergy}). A sharp jump in the energy of the state at ${\bf{k}} = \left( {\pi ,\pi } \right)$ is observed at a concentration of $x=0.234$. This feature indicates a possible relationship between a sharp increase in the $s^*$-wave component of the superconducting gap and a decrease in the depth of the inner small electron pocket since the maximum increase in the $s^*$-wave component is observed at the same concentration $x=0.234$. The inner electron pocket shallowing manifests itself in the density of states as the appearance of additional states near the chemical potential (Fig.~\ref{fig:band_str}c,d, panel on the left).

Although the pairing involves states with all momenta ${\bf{k}}$ the largest contribution to the superconducting gap is made by the states on the Fermi contours. The ${\bf{k}}$-dependence of the superconducting gap for the states of the Fermi contours is seen from the positions of the BCS spectral function peaks at ${{\bf{k}}_F}$ when passing from points $A$ and $C$ of the nodal direction to points $B$ and $D$ of the antinodal direction (Fig.~\ref{fig:Fermi_cont}b, inset of Fig.~\ref{fig:spect_func}b).  The superconducting gap is close to zero at points $A$ and $C$ (Fig.~\ref{fig:spect_func}a). The gap at point $A$ of the large outer Fermi contour is smaller than at point $C$ for the smaller inner Fermi contour (Fig.~\ref{fig:spect_func}a) since point $A$ is closer to the line of zeros of the gap (Fig.~\ref{fig:Fermi_cont}b, red dashed line) than point $C$. The latter fact is caused by the stronger deviation of the gap nodes from the nodal direction in the ${\bf{k}}$-region where the inner pocket is located. The superconducting gap at point $A - \left( {0.44\pi ,0.44\pi } \right)$ is $0.2$ eV larger than its value at point $B- \left( {\pi ,0.32\pi } \right)$ (Fig.~\ref{fig:spect_func}a). Similar growth of the superconducting gap is observed when passing along the inner Fermi contour from the nodal point $C-\left( {0.74\pi ,0.74\pi } \right)$  to the antinodal point $D- \left( {\pi ,0.6\pi } \right)$ (Fig.~\ref{fig:spect_func}b), but the gap increases by only $0.07$ eV in this case. Thus superconducting gap for the states of the outer contour is more anisotropic than for the states of the inner contour.

The concentration dependence of the superconducting gap has a different character in different regions of the momentum space. It can be seen that dependence of the superconducting gap on the Fermi angle $\Theta $ (inset in Figs.~\ref{fig:gap_profile}a,b) along the outer Fermi contour at $x=0.21$ and $0.22$ (Fig.~\ref{fig:gap_profile}a, black and red lines) is almost described by the $d$-wave (Fig.~\ref{fig:gap_profile}a, magenta dotted line) except for slight deviations due to the orthorhombic distortion. The gap decreases on most of the outer contour with increasing doping while it increases slightly at points near the nodal direction (Figs.~\ref{fig:spect_func}a,c) due to the fact that the minimum of the gap moves away from this direction. The deviation of the superconducting gap minimum position from $\Theta  = 45^\circ $  and the difference in the gap maxima at the points of directions $\left( {0,\pi } \right) - \left( {\pi ,\pi } \right)$ ($\Theta  = 0^\circ $) and $\left( {\pi ,0} \right) - \left( {\pi ,\pi } \right)$ ($\Theta  = 90^\circ $) becomes noticeable at $x = 0.23$ and $x = 0.24$ (Fig.~\ref{fig:gap_profile}a, blue and green lines). Remarkable different behavior with doping is observed for the profile of the superconducting gap along the inner Fermi contour. The zero of the gap is significantly displaced from the nodal direction $\left( {0,0} \right) - \left( {\pi ,\pi } \right)$ ($\Theta  = 45^\circ $) even at $x = 0.22$. The gap is absent on the inner contour at $x=0.23$ since the line of zeros ceases to intersect the contour (Fig.~\ref{fig:Fermi_cont}c), and therefore the superconducting gap profile differs significantly from the $d$-wave gap. An even greater deviation from the $d$-wave form is observed at $x=0.23$ and $x=0.24$: the gap minimum is in the direction $\left( {0,\pi } \right) - \left( {\pi ,\pi } \right)$, and the maximum is in the direction $\left( {\pi ,0} \right) - \left( {\pi ,\pi } \right)$ (Fig.~\ref{fig:gap_profile}b). There is a significant increase in the superconducting gap at ${{\bf{k}}_F}$ along most of the inner contour, the energy shift of the spectral function at ${\bf{k}}$-point $C$ peak is $0.035$ eV with an increase in doping from $0.21$ to $0.23$ (Figs.~\ref{fig:spect_func}b,d). The spectral function peak at ${\bf{k}}$-point $D$ is hardly shifted with doping.

The main features of the concentration dependence of the electronic structure are the shallowing of the inner electron pocket around $\left( {\pi ,\pi } \right)$ and the growth of spectral weight of the states in it. The other important result which is based on the study of the profile of the superconducting gap along the outer and inner Fermi contours is the difference in the ${\bf{k}}$-dependence of the gap in various regions of the Brillouin zone and the change in ${\bf{k}}$-dependence upon doping. To understand whether there is a relationship between these two effects, it is necessary to analyze the evolution of partial contributions to the superconducting gap of paired states with different wave vectors depending on doping.
\section{\label{sec:partcontr}The analysis of the momentum-dependent structure of the partial contributions of paired states to the superconducting gap at different dopings}

\begin{figure*}
\includegraphics[width=0.45\linewidth]{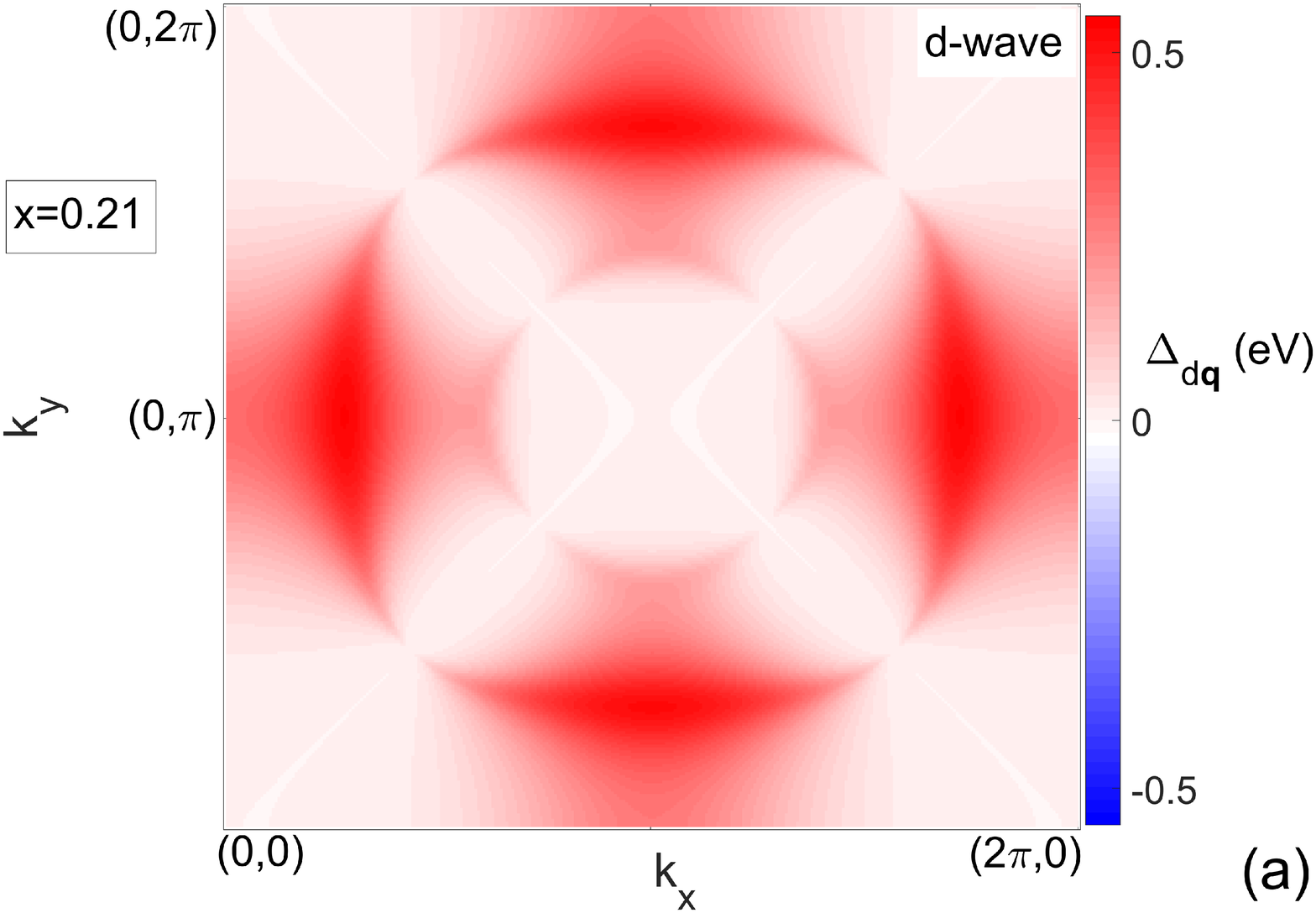}
\includegraphics[width=0.45\linewidth]{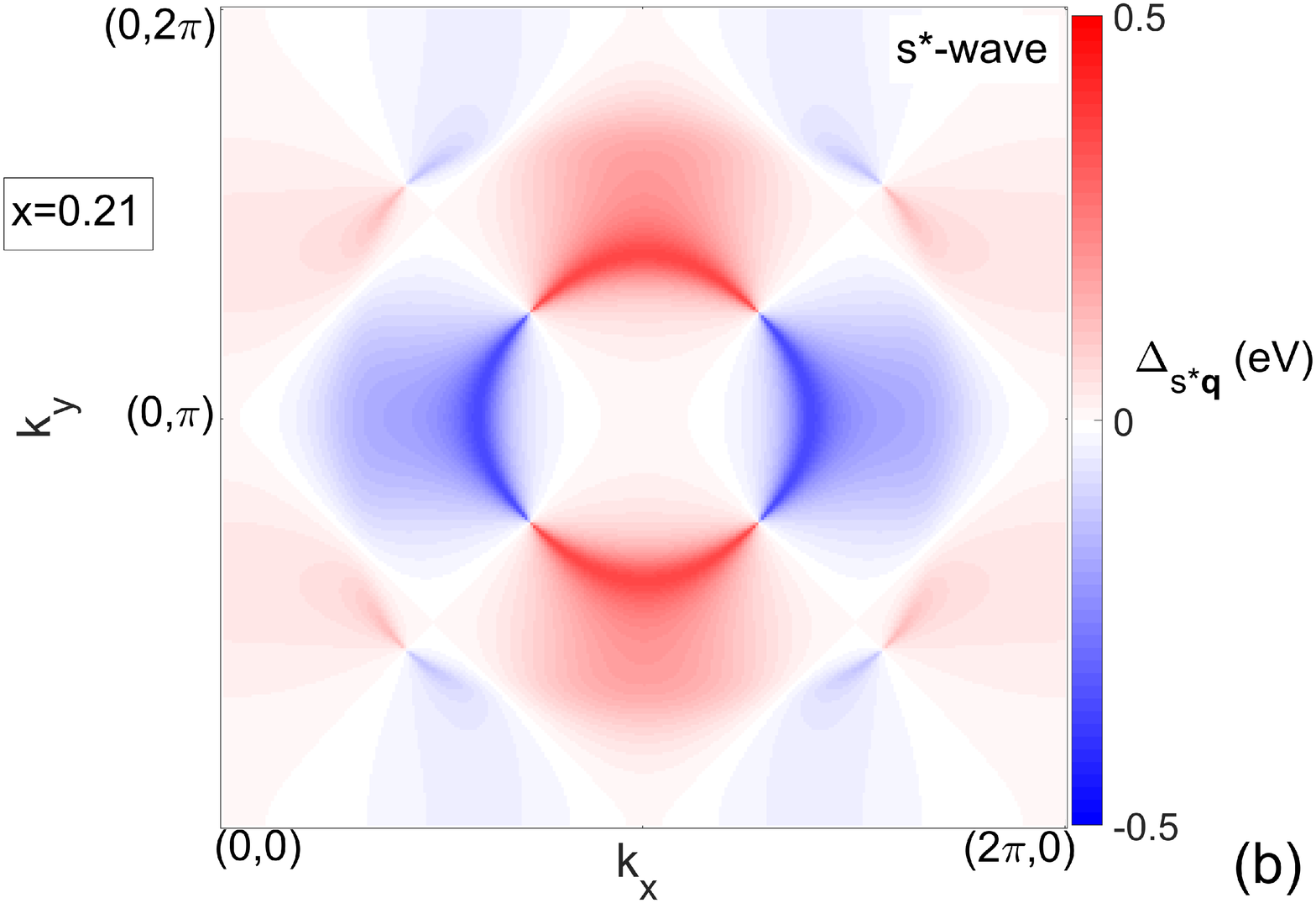}
\caption{\label{fig:partcontr_d_s} Maps of the partial contributions ${\Delta _{d{\bf{q}}}}$ (a) and ${\Delta _{s*{\bf{q}}}}$ (b) of paired states with wave vectors ${\bf{q}}$ and $ - {\bf{q}}$ to the $d$-wave and extended $s^*$-wave components of superconducting gap, respectively, at doping $x = 0.21$. Red (blue) color with different intensity denotes positive (negative) contributions ${\Delta _{d{\bf{q}}}}$ and ${\Delta _{s*{\bf{q}}}}$ with different magnitudes.}
\end{figure*}

\begin{figure*}
\includegraphics[width=0.6\linewidth]{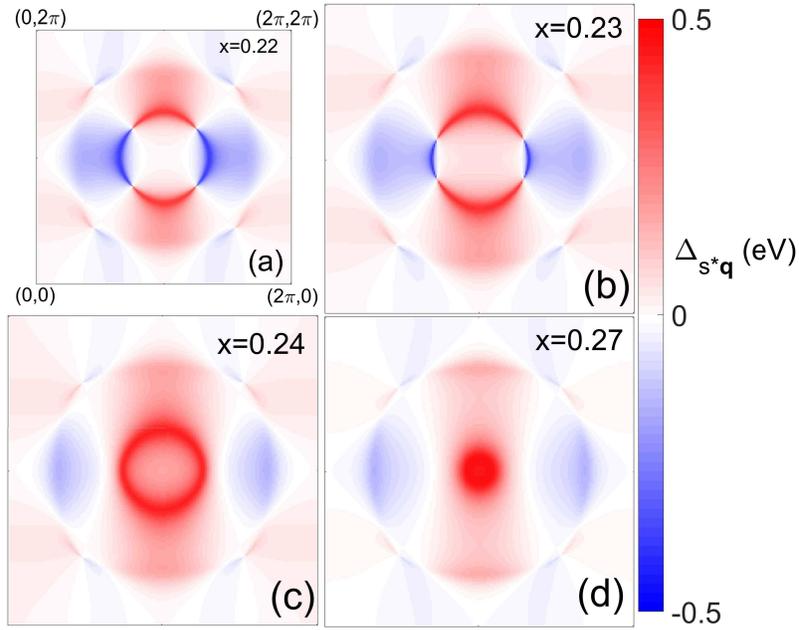}
\caption{\label{fig:partcontr_s} The doping dependence of the surface of the partial contributions ${\Delta _{s*{\bf{q}}}}$ of pair states with wave vectors ${\bf{q}}$ and $ - {\bf{q}}$ to the extended $s^*$-wave component of the superconducting gap. Red (blue) color with different intensity denotes positive (negative) contributions ${\Delta _{s*{\bf{q}}}}$ with different magnitudes.}
\end{figure*}

\begin{figure*}
\includegraphics[width=0.45\linewidth]{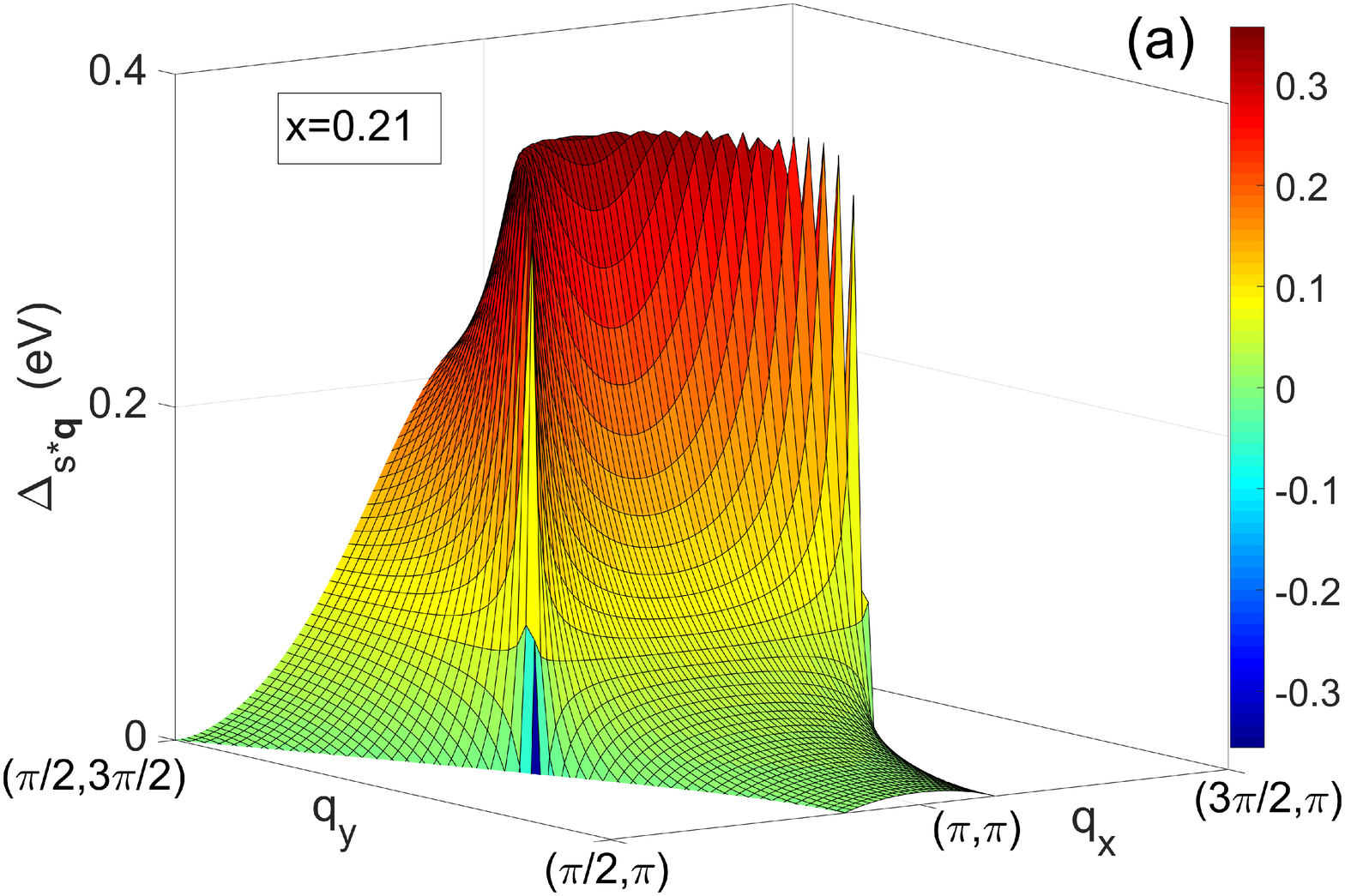}
\includegraphics[width=0.45\linewidth]{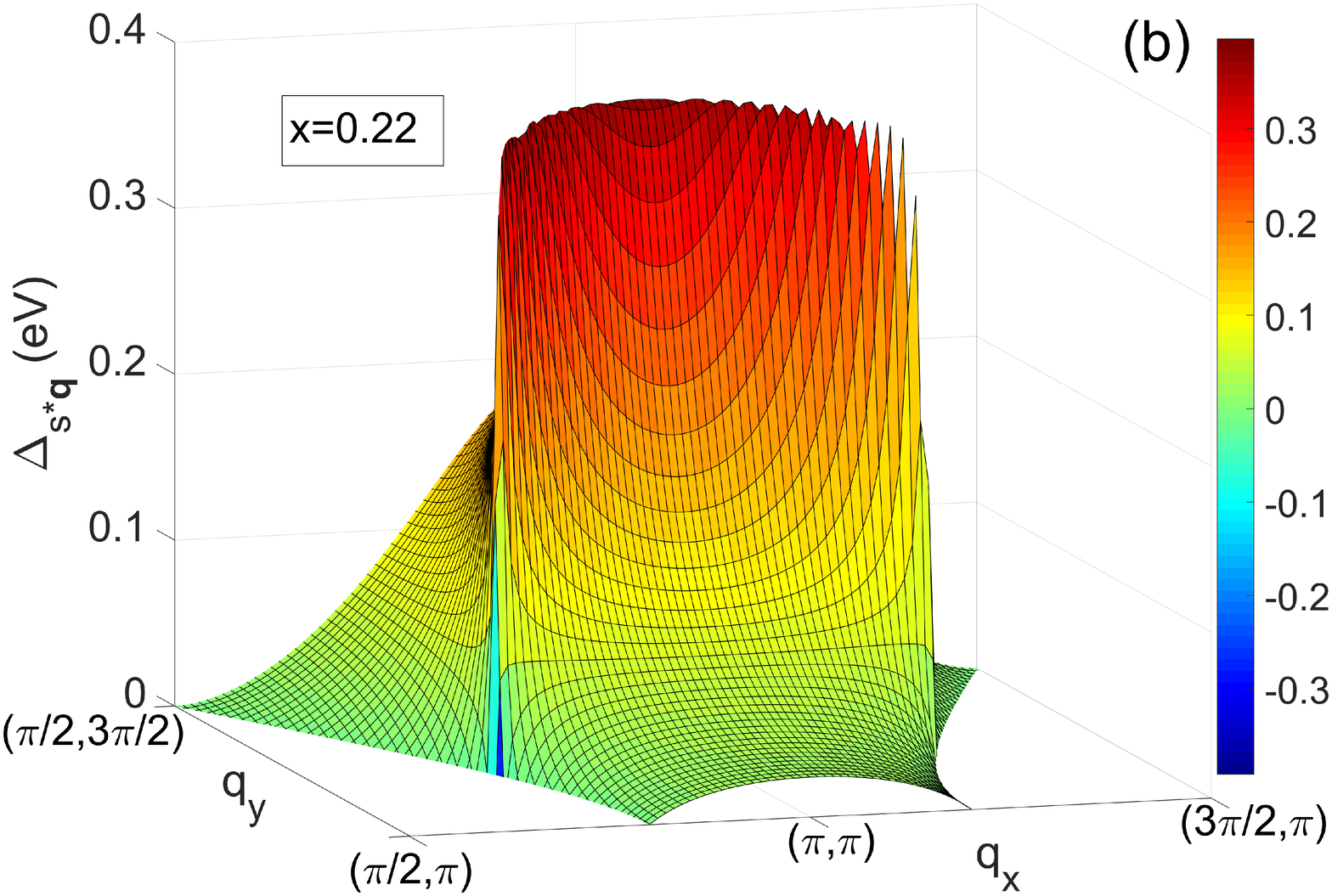}
\includegraphics[width=0.45\linewidth]{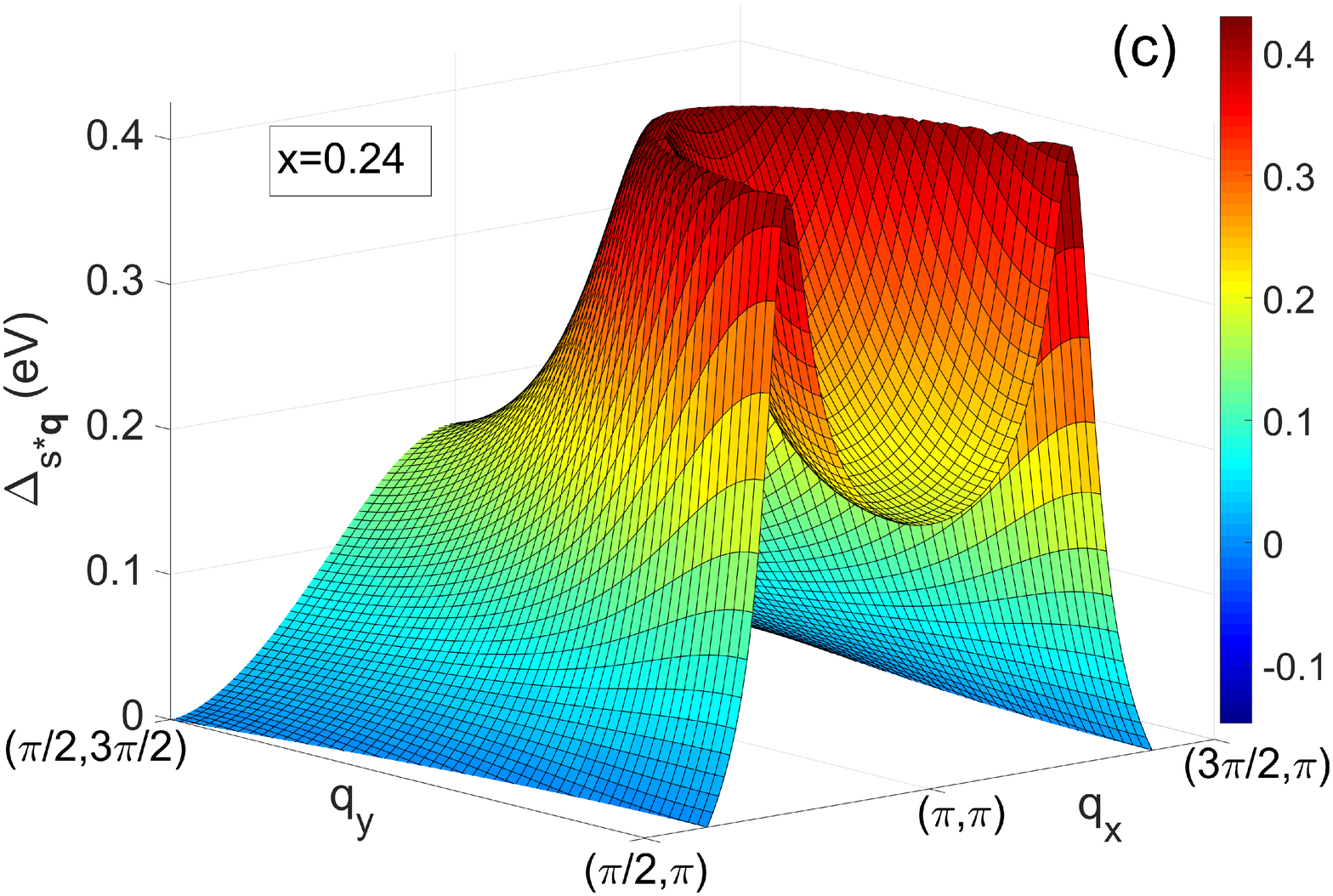}
\includegraphics[width=0.45\linewidth]{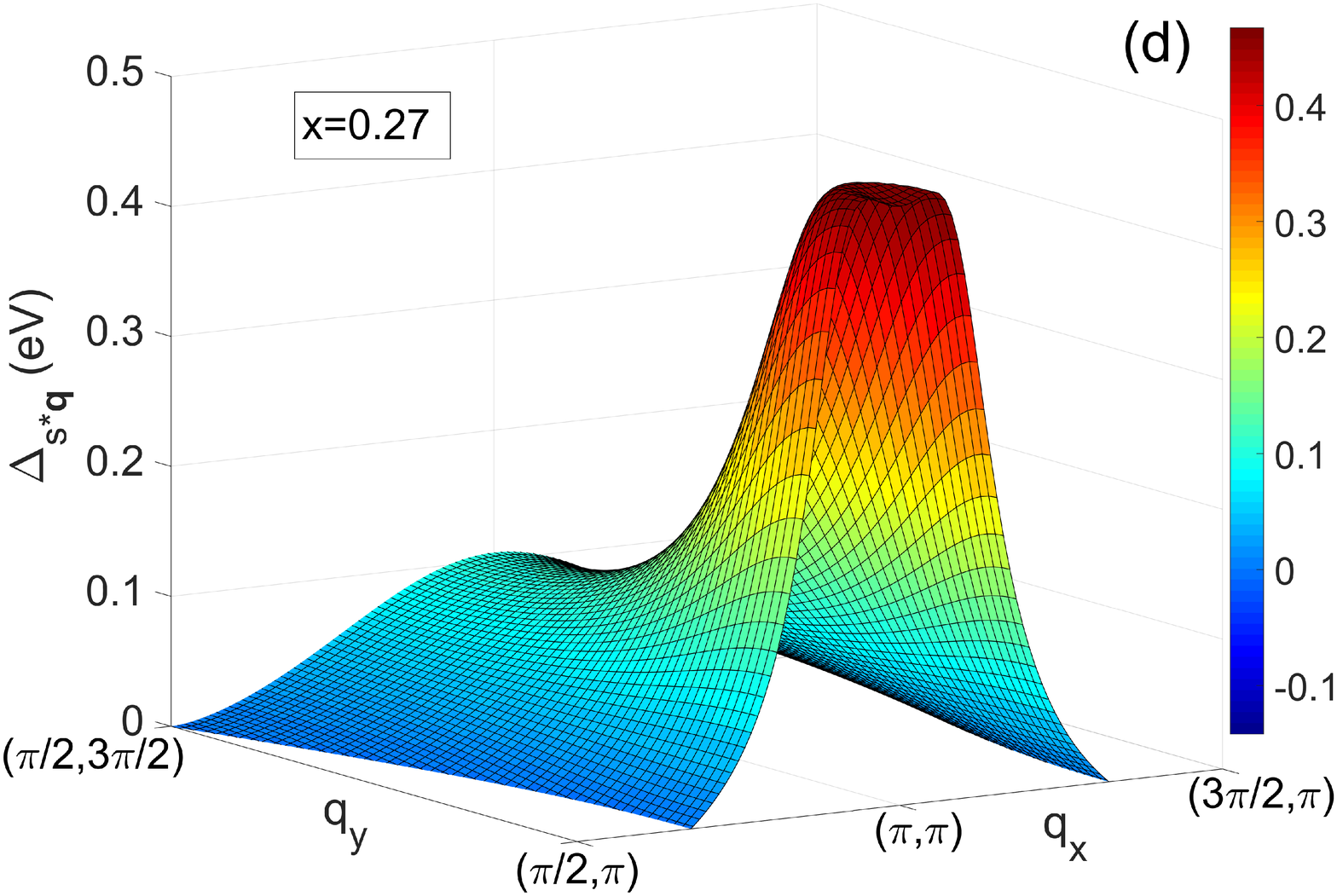}
\caption{\label{fig:cuts_s_gap} Cuts of the surfaces of the partial contributions ${\Delta _{s*{\bf{q}}}}$ at different doping. It is seen the growth of contributions from states of the inner electron pocket in the vicinity of a point $\left( {\pi ,\pi } \right)$ to the $s^*$-wave component of the superconducting gap with doping.}
\end{figure*}

\begin{figure*}
\includegraphics[width=0.45\linewidth]{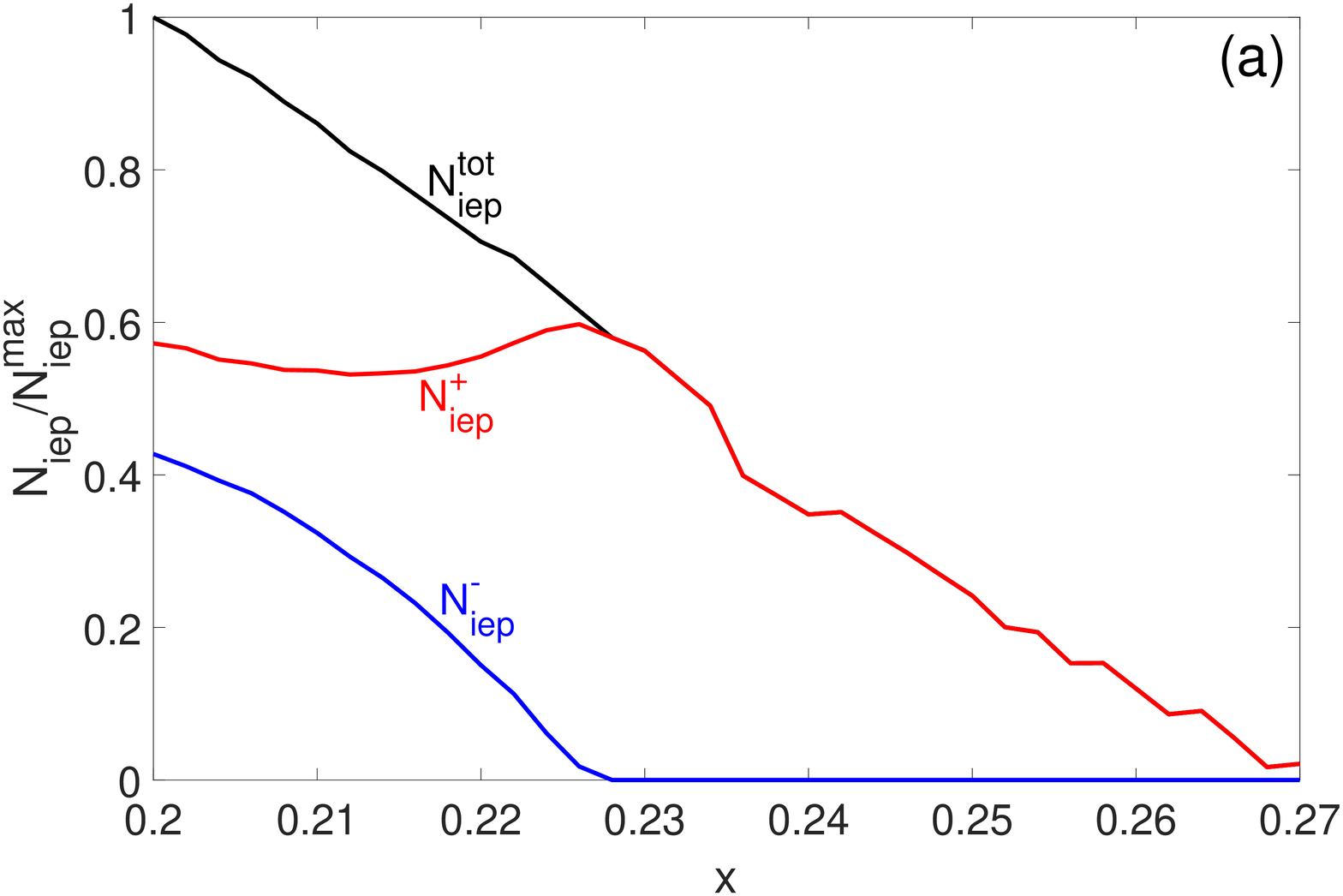}
\includegraphics[width=0.45\linewidth]{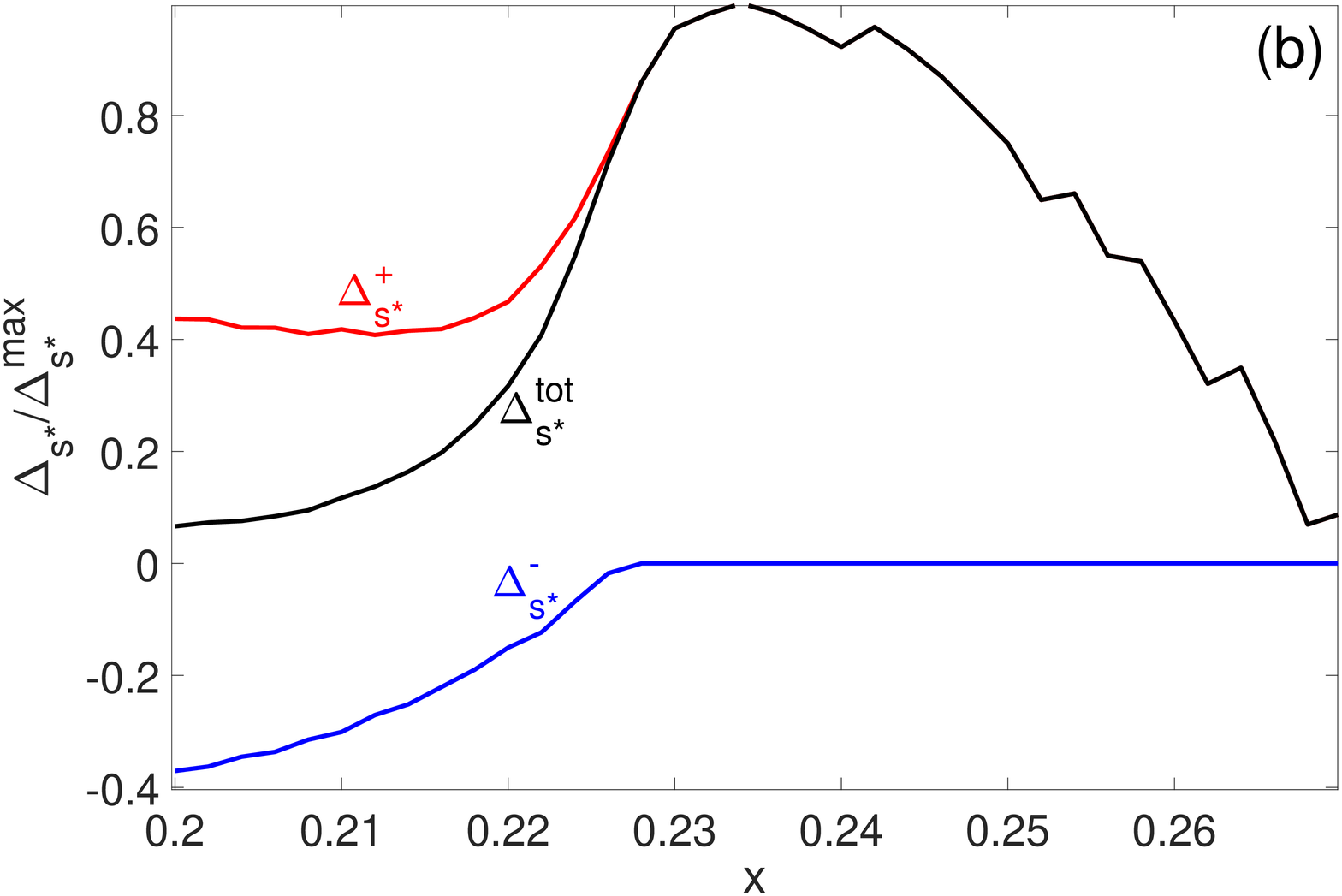}
\caption{\label{fig:Ncontr} The number of positive $N_{iep}^ + $ (red line), negative $N_{iep}^ - $ (blue line), and total contributions $N_{iep}^{tot}$ (black line) of the inner electron pocket states to the $s^*$-wave gap. (b) The sum positive contribution $\Delta _{s*}^ + $ (red line), the sum negative contribution $\Delta _{s*}^ - $ (blue line), and total contribution ${\Delta _{s*}}$ (black line) of the inner electron pocket states to the $s^*$-wave gap.}
\end{figure*}

The structure of partial contributions of paired states with certain wave vectors ${\bf{q}}$ and $ - {\bf{q}}$ to the superconducting gap components at different doping will be used as a tool to analyze the concentration dependence of the ratio of these components. The partial contribution of paired states with momenta ${\bf{q}}$ and $ - {\bf{q}}$ to components ${\Delta _d}$ and ${\Delta _{s*}}$ is defined by the sum elements ${\Delta _{d{\bf{q}}}}$ of Eq.~(\ref{eq:gap_d}) and ${\Delta _{s*{\bf{q}}}}$ of Eq.~(\ref{eq:gap_s}), respectively. The surfaces of the contributions ${\Delta _{d{\bf{q}}}}$ and ${\Delta _{s*{\bf{q}}}}$ in the first Brillouin zone are shown in Fig.~\ref{fig:partcontr_d_s}. The main contributions to the $d$-wave component arise from the outer Fermi contour and states in its vicinity (Fig.~\ref{fig:partcontr_d_s}a), the contributions from the inner Fermi contour are much smaller. The main contributions to the $s^*$-wave component come from the inner contour (Fig.~\ref{fig:partcontr_d_s}b). It can be seen that the contributions to ${\Delta _d}$ from states with wave vectors over the entire Brillouin zone have the same (positive) sign (Fig.~\ref{fig:partcontr_d_s}a). The contributions to ${\Delta _{s*}}$ from regions of states with ${\bf{q}}$ along direction $\left( {\pi ,0} \right) - \left( {\pi ,2\pi } \right)$ and along direction $\left( {0,\pi } \right) - \left( {2\pi ,\pi } \right)$ including arcs of the inner contour have different signs (Fig.~\ref{fig:partcontr_d_s}b). The regions of ${\Delta _{s*{\bf{q}}}}$ with different signs are separated by the nodes of the superconducting gap. The size of the sections with positive and negative contributions to $s^*$-wave component ${\Delta _{s*}}$ is the same in a system without orthorhombic distortion with a superconducting gap of pure $d$-wave symmetry, the electronic structure of such tetragonal structure is symmetric with respect to reflection in planes containing a line of zeros and perpendicular to the ${k_x} - {k_y}$ plane. The contributions ${\Delta _{s*{\bf{q}}}}$ in the tetragonal structure are canceled when summing over all wave vectors (over regions with both positive and negative signs) and ${\Delta _{s*}}$ becomes zero. In a system with orthorhombic distortion, the electronic structure has no reflection planes in the directions $\left( {0,0} \right) - \left( {2\pi ,2\pi } \right)$ and $\left( {2\pi ,0} \right) - \left( {0,2\pi } \right)$ of momentum space. The line of gap zeros turns into curves when adding the $s^*$-wave component (Figs.~\ref{fig:Fermi_cont}a-d). There is a difference in total contributions to ${\Delta _{s*{\bf{q}}}}$ from regions with positive and negative signs. Therefore the existence of pure $d$-wave symmetry of the superconducting gap is impossible in the system with orthorhombic distortion, a nonzero $s^*$-wave component appears.

In the concentration range from $x=0.21$ to $x=0.27$, the inner electron pocket reaches its largest size at doping $x=0.21$. The gap has predominantly $d$-wave symmetry, the positive and negative contributions to the ${\Delta _{s*}}$ are almost completely compensated leaving only a small value of the $s^*$-wave component. The most noticeable uncompensated contribution to admixing of the $s^*$-wave component to the $d$-wave component is made by the states with momenta in the vicinity of the ${\bf{k}}$-point $\left( {\pi ,\pi } \right)$ and along the direction $\left( {\pi ,0} \right) - \left( {\pi ,2\pi } \right)$ between the separated lines of zeros of the superconducting gap (Figs.~\ref{fig:partcontr_s}a,b). The $s^*$-wave superconducting gap at the ${\bf{k}}$-point $\left( {\pi ,\pi } \right)$ has a nonzero value (Fig.~\ref{fig:cuts_s_gap}a) at $x=0.21$ in contrast to pure $d$-symmetry. Figs.~\ref{fig:cuts_s_gap}a-d show cuts of the surfaces of the ${\bf{q}}$-dependent partial contributions ${\Delta _{s*{\bf{q}}}}$ at different concentrations, these cuts serve to demonstrate the growth of positive contributions in the depth of the inner electron pocket with doping. The number of pair states inside the inner pocket positively ($N_{iep}^ + $) and negatively ($N_{iep}^ - $) contributing to ${\Delta _{s*}}$ at $x<0.215$ both decreases together with total contraction of this pocket (Fig.~\ref{fig:Ncontr}a). When the doping rises above $x=0.215$ the depth of the inner electron pocket becomes small enough for the pairing mechanism to more intensively activate the states on its entire electron pocket surface and not only on the contour. These states inside the inner electron pocket are involved in the pairing process with $s^*$-wave symmetry since they are in the region of wave vectors in which the $s^*$-wave gap has maximum values and the $d$-wave gap has minimum values. The increase in the magnitude of the contributions ${\Delta _{s*{\bf{q}}}}$ from the states inside the inner pocket at $x=0.22$ is seen in Fig.~\ref{fig:cuts_s_gap}b; the contributions in the depth of the inner pocket have a positive sign and are not compensated by negative contributions. The increase in the number of positive contributions $N_{iep}^ + $ is also seen in Fig.~\ref{fig:Ncontr}a (red line) as hump at $x=0.226$. The predominance of positive contributions is also achieved because the arcs of the inner contour with states giving positive contributions become longer than the arcs of negative signs (Fig.~\ref{fig:partcontr_s}b). The difference in the length of the Fermi contour arcs which give contributions of different signs to the gap is the result of a change in the shape of the gap nodes and a reduction in the size of the pocket. The negative contributions disappear at $x=0.228$ (Fig.~\ref{fig:Ncontr}a,b, blue line) when the inner electron pocket is so small that it fits entirely in the region between the nodes (Fig.~\ref{fig:Fermi_cont}c) and the total gap is formed by the only positive contributions (Fig.~\ref{fig:partcontr_s}c,d;Fig.~\ref{fig:Ncontr}a, red line). The growth in the number of positive contributions ends quickly but the magnitude of positive contributions $\Delta _{s*{\bf{q}}}^ + $ continues to grow (Fig.~\ref{fig:Ncontr}b, red line). As a result, $\Delta _{s*}^ + $ reaches the maximum at $x=0.234$, it defines the maximum of the $s^*$-wave component (Fig.~\ref{fig:Ncontr}b). Even more noticeable contribution ${\Delta _{s*{\bf{q}}}}$ around point $\left( {\pi ,\pi } \right)$ at $x>0.234$ is made by states in the depth of the pocket (Fig.~\ref{fig:cuts_s_gap}c,d) but the number of contributions $N_{iep}^ + $ and total positive contribution $\Delta _{s*}^ + $ decrease because of the contraction of the inner pocket.

Thus two main competing processes control the value of the $s^*$-wave components with increasing hole concentration. First, the shallowing inner pocket that increases the number of states involved in pairing and amplitude of their pairing results in the growth of the $s^*$-wave superconducting gap magnitude. Second, the decrease in the total number of the pair states in the inner electron pocket suppresses the $s^*$-wave component magnitude. Competition of these two tendencies leads to the maximum of the $s^*$-wave component approximately at $x=0.234$.
\section{\label{sec:Concl}Conclusion}
In this work, the enhanced extended $s^*$-wave component of the superconducting gap in the region of strong doping of cuprates with orthorhombic distortion is obtained and the reasons for this effect are investigated. The extended $s^*$-wave symmetry is mixed with the $d$-wave symmetry because of orthorhombicity but its fraction is small in the region of weak and optimal doping. In the region of strong doping, the Fermi contour is formed by the large outer and small inner contours around the point $\left( {\pi ,\pi } \right)$. The inner electron pocket states are responsible for the enhanced $s^*$-wave component since they are located in the region of momentum space in which the extended $s^*$-wave gap has a maximum value and the $d$-wave gap has a minimum value. Two opposite processes occur with doping. On the one hand, the inner electron pocket decreases, that is, the number of states in this pocket decreases. On the other hand, the pocket becomes shallower, the states in the depth of this pocket become higher in energy and rise to the level of the chemical potential. The pocket depth becomes so small at a certain concentration of doped holes that states in the depth of the pocket are involved in pairing. These states make a large contribution to the $s^*$-wave symmetry gap. The total number of states decreases with further doping and the total contribution to the $s^*$-wave component decreases although the magnitude of partial contributions in the inner pocket continues to grow. Thus it turns out that the $s^*$-wave component fraction under certain conditions depends on the electronic structure much more strongly than on the degree of orthorhombicity. Based on the study a general conclusion can be drawn regarding the ratio of the several superconducting gap components: the condition for the significant growth of a certain component of the superconducting gap is the presence of shallow and large enough in area pockets in the region of the momentum space where this component has a significant value and the competing components are minimal. Observation of the mixing of the $s^*$-wave component in various experiments can be explained by the presence of elements of the electronic structure (such as shallow electron pockets) with the states which significantly contribute to the superconducting gap.

\begin{acknowledgments}
We would like to especially thank A. Bianconi for new ideas, useful discussions, and suggestions. The reported study was funded by Russian Foundation for Basic Research, Government of Krasnoyarsk Territory and Krasnoyarsk Regional Fund of Science according to the research project "Studies of superexchange and electron-phonon interactions in correlated systems as a basis for searching for promising functional materials" No. 20-42-240016
\end{acknowledgments}

\appendix
\section{\label{app:fillnumb}Filling numbers of local eigenstates}
Filling numbers of local eigenstates $\left\langle {{X^{pp}}} \right\rangle $ are determined self-consistently from the condition of completeness $\sum\limits_p {X_f^{pp} = 1} $, the chemical potential equation:
\begin{equation}
n = 1 + x = 0 \cdot \left\langle {{X^{0l,0l}}} \right\rangle  + \sum\limits_\sigma  {1 \cdot \left\langle {{X^{\sigma \sigma }}} \right\rangle }  + 2 \cdot \left\langle {{X^{SS}}} \right\rangle
\label{eq:chempoteq} 
\end{equation}
(here $n = 1 + x$ is the hole concentration for La$_{2-x}$Sr$_x$CuO$_4$) and relation $\left\langle {{X^{SS}}} \right\rangle  = \left\langle {{X^{00}}} \right\rangle  + x$. Filling numbers for the zero-, single- and two-hole states $\left\langle {{X^{00}}} \right\rangle $, $\left\langle {{X^{\sigma \sigma }}} \right\rangle $, $\left\langle {{X^{SS}}} \right\rangle $ are defined by the formulas
\begin{widetext}
\begin{eqnarray}
\left\langle {{X^{00}}} \right\rangle & = & \left( { - \frac{1}{\pi }} \right)\frac{1}{N}\sum\limits_{\bf{k}} {\int_{ - \infty }^\infty  {\frac{1}{{\exp \left( {{{\left( {\omega  - \mu } \right)} \mathord{\left/
 {\vphantom {{\left( {\omega  - \mu } \right)} {kT}}} \right.
 \kern-\nulldelimiterspace} {kT}}} \right) + 1}}{\mathop{\rm Im}\nolimits} {{\left\langle {\left\langle {{X_{\bf{k}}^{0\sigma }}}
 \mathrel{\left | {\vphantom {{X_{\bf{k}}^{0\sigma }} {X_{\bf{k}}^{\sigma 0}}}}
 \right. \kern-\nulldelimiterspace}
 {{X_{\bf{k}}^{\sigma 0}}} \right\rangle } \right\rangle }_{\omega  + i\delta }}} d\omega }, \\
\left\langle {{X^{\sigma \sigma }}} \right\rangle & = & \left( { - \frac{1}{\pi }} \right)\frac{1}{N}\sum\limits_{\bf{k}} {\int_{ - \infty }^\infty  {\frac{1}{{\exp \left( {{{ - \left( {\omega  - \mu } \right)} \mathord{\left/
 {\vphantom {{ - \left( {\omega  - \mu } \right)} {kT}}} \right.
 \kern-\nulldelimiterspace} {kT}}} \right) + 1}}{\mathop{\rm Im}\nolimits} {{\left\langle {\left\langle {{X_{\bf{k}}^{0\sigma }}}
 \mathrel{\left | {\vphantom {{X_{\bf{k}}^{0\sigma }} {X_{\bf{k}}^{\sigma 0}}}}
 \right. \kern-\nulldelimiterspace}
 {{X_{\bf{k}}^{\sigma 0}}} \right\rangle } \right\rangle }_{\omega  + i\delta }}} d\omega }, \\
\left\langle {{X^{SS}}} \right\rangle & = & \left( { - \frac{1}{\pi }} \right)\frac{1}{N}\sum\limits_{\bf{k}} {\int_{ - \infty }^\infty  {\frac{1}{{\exp \left( { - {{\left( {\omega  - \mu } \right)} \mathord{\left/
 {\vphantom {{\left( {\omega  - \mu } \right)} {kT}}} \right.
 \kern-\nulldelimiterspace} {kT}}} \right) + 1}}{\mathop{\rm Im}\nolimits} {{\left\langle {\left\langle {{X_{\bf{k}}^{\sigma S}}}
 \mathrel{\left | {\vphantom {{X_{\bf{k}}^{\sigma S}} {X_{\bf{k}}^{S\sigma }}}}
 \right. \kern-\nulldelimiterspace}
 {{X_{\bf{k}}^{S\sigma }}} \right\rangle } \right\rangle }_{\omega  + i\delta }}} d\omega }.
\label{eq:fillnumbers}
\end{eqnarray}
\end{widetext}
Note filling numbers in zero-hole and two-hole sectors $\left\langle {{X^{00}}} \right\rangle $ and $\left\langle {{X^{SS}}} \right\rangle $ are nonzero even in the undoped compound ($x = 0$). This fact is caused by the hybridization of quasiparticle excitations between zero- and single-hole states and excitations between single- and two-hole states.
\section{\label{app:kinpar}Kinematic parameters}
Terms $T_{\bf{k}}^{\alpha \beta }$ in generalized mean-field approximation contain hoppings, kinematic, and spin-spin correlation functions: 
\begin{widetext}
\begin{eqnarray}
\bar T_{\bf{k}}^{11}& = & F\left( {0\bar \sigma } \right)\bar t_{\bf{k}}^{11} + \frac{1}{N}\frac{1}{{F\left( {0\bar \sigma } \right)}}\sum\limits_{\bf{q}} {\left( {t_{{\bf{k}} - {\bf{q}}}^{11} + \frac{1}{2}\bar t_{{\bf{k}} - {\bf{q}}}^{11}} \right){C_{\bf{q}}}}  + \frac{1}{{F\left( {0\bar \sigma } \right)}}{a^{11}}, \\\nonumber
\bar T_{\bf{k}}^{12} & = & F\left( {0\bar \sigma } \right)\bar t_{\bf{k}}^{12} + \frac{1}{N}\frac{1}{{F\left( {\sigma S} \right)}}\sum\limits_{\bf{q}} {\left( {t_{{\bf{k}} - {\bf{q}}}^{12} - \frac{1}{2}\bar t_{{\bf{k}} - {\bf{q}}}^{12}} \right){C_{\bf{q}}}}  + \frac{1}{{F\left( {\sigma S} \right)}}{a^{12}}, \\\nonumber
\bar T_{\bf{k}}^{21} & = & F\left( {\sigma S} \right)\bar t_{\bf{k}}^{21} + \frac{1}{N}\frac{1}{{F\left( {0\bar \sigma } \right)}}\sum\limits_{\bf{q}} {\left( {t_{{\bf{k}} - {\bf{q}}}^{21} - \frac{1}{2}\bar t_{{\bf{k}} - {\bf{q}}}^{21}} \right){C_{\bf{q}}}}  + \frac{1}{{F\left( {0\bar \sigma } \right)}}{a^{21}}, \\\nonumber
\bar T_{\bf{k}}^{22} & = & F\left( {\sigma S} \right)\bar t_{\bf{k}}^{22} + \frac{1}{N}\frac{1}{{F\left( {\sigma S} \right)}}\sum\limits_{\bf{q}} {\left( {t_{{\bf{k}} - {\bf{q}}}^{22} + \frac{1}{2}\bar t_{{\bf{k}} - {\bf{q}}}^{22}} \right){C_{\bf{q}}}}  + \frac{1}{{F\left( {\sigma S} \right)}}{a^{22}}, \\\nonumber
T_{\bf{k}}^{11} & = & F\left( {0\sigma } \right)t_{\bf{k}}^{11} + \frac{1}{N}\frac{1}{{F\left( {0\sigma } \right)}}\sum\limits_{\bf{q}} {\left( {\bar t_{{\bf{k}} - {\bf{q}}}^{11} + \frac{1}{2}t_{{\bf{k}} - {\bf{q}}}^{11}} \right){C_{\bf{q}}}}  + \frac{1}{{F\left( {0\sigma } \right)}}{{\bar a}^{11}}, \\\nonumber
T_{\bf{k}}^{12} & = & F\left( {\bar \sigma S} \right)t_{\bf{k}}^{12} + \frac{1}{N}\frac{1}{{F\left( {0\sigma } \right)}}\sum\limits_{\bf{q}} {\left( {\bar t_{{\bf{k}} - {\bf{q}}}^{12} - \frac{1}{2}t_{{\bf{k}} - {\bf{q}}}^{12}} \right){C_{\bf{q}}}}  + \frac{1}{{F\left( {0\sigma } \right)}}{{\bar a}^{21}}, \\\nonumber
T_{\bf{k}}^{21} & = & F\left( {0\sigma } \right)t_{\bf{k}}^{21} + \frac{1}{N}\frac{1}{{F\left( {\bar \sigma S} \right)}}\sum\limits_{\bf{q}} {\left( {\bar t_{{\bf{k}} - {\bf{q}}}^{21} - \frac{1}{2}t_{{\bf{k}} - {\bf{q}}}^{21}} \right){C_{\bf{q}}}}  + \frac{1}{{F\left( {\bar \sigma S} \right)}}{{\bar a}^{12}}, \\\nonumber
T_{\bf{k}}^{22} & = & F\left( {\bar \sigma S} \right)t_{\bf{k}}^{22} + \frac{1}{N}\frac{1}{{F\left( {\bar \sigma S} \right)}}\sum\limits_{\bf{q}} {\left( {\bar t_{{\bf{k}} - {\bf{q}}}^{22} + \frac{1}{2}t_{{\bf{k}} - {\bf{q}}}^{22}} \right){C_{\bf{q}}}}  + \frac{1}{{F\left( {\bar \sigma S} \right)}}{{\bar a}^{22}}. 
\label{eq:kinenergy}
\end{eqnarray}
\end{widetext}
In these formulas $t_{\bf{k}}^{\alpha \beta } = \sum\limits_{\lambda \lambda '} {\gamma _\lambda ^ * \left( \alpha  \right)} {\gamma _{\lambda '}}\left( \beta  \right){t_{\lambda \lambda '}}\left( {\bf{k}} \right)$ are intra- and interband hoppings of the Hubbard fermions $\alpha $ and $\beta $ expressed in terms of electron hoppings ${t_{\lambda \lambda '}}\left( {\bf{k}} \right)$ between orbitals $\lambda $ and $\lambda '$. The quasiparticle excitation $\alpha $ is defined as a transition between certain multielectron CuO$_6$ cluster eigenstates $\left| p \right\rangle $ and $\left| q \right\rangle $ with the number of particles differing by one, ${\gamma _\lambda }\left( \alpha  \right) = {\gamma _\lambda }\left( {pq} \right) = \left\langle p \right|{a_\lambda }\left| q \right\rangle $ is the amplitude of such transition, where ${a_\lambda }$ is the annihilation operator of electron on orbital $\lambda $. An explicit form of hoppings $t_{\bf{k}}^{\alpha \beta }$ in the layer of CuO$_6$ octahedra with orthorhombic distortion is given in the work~\cite{Makarov21}. $t_{\bf{k}}^{11}$ is hopping integral of quasiparticles with spin projection $\sigma  = {1 \mathord{\left/
 {\vphantom {1 2}} \right.
 \kern-\nulldelimiterspace} 2}$ inside conductivity band, $t_{\bf{k}}^{22}$ is hopping integral inside valence band, $t_{\bf{k}}^{12}$, $t_{\bf{k}}^{21}$ are hoppings between conductivity and valence bands. $\bar t_{\bf{k}}^{11},\bar t_{\bf{k}}^{12},\bar t_{\bf{k}}^{21},\bar t_{\bf{k}}^{22}$ are the same hopping integrals of quasiparticles with spin projection $\sigma  = {{ - 1} \mathord{\left/
 {\vphantom {{ - 1} 2}} \right.
 \kern-\nulldelimiterspace} 2}$. ${C_{\bf{q}}} = \sum\limits_{fg} {{C_{fg}}{e^{i\left( {f - g} \right){\bf{q}}}}} $ is the static spin-spin correlation function, ${C_{fg}} = \left\langle {X_f^{\sigma \bar \sigma }X_g^{\bar \sigma \sigma }} \right\rangle $. Values of spin-spin correlation functions for different hole concentrations are taken from~\cite{Korshunov07}. Terms $a^{11}$, $a^{12}$, $a^{21}$, $a^{22}$ containing kinematic correlation functions have the form:
 
\begin{eqnarray}
&&a^{11} = \sum\limits_g {\left( {t_{gj}^{22}K_{gj}^{22} - t_{gj}^{11}K_{gj}^{11}} \right)}, \\\nonumber
&&a^{12} = \sum\limits_g {\left( {t_{jg}^{11}K_{jg}^{21} - t_{gj}^{22}K_{gj}^{21} + t_{jg}^{12}K_{jg}^{22} - t_{gj}^{12}K_{gj}^{11}} \right)}, \\\nonumber
&&a^{21} = \sum\limits_g {\left( {t_{gj}^{11}K_{gj}^{12} - t_{jg}^{22}K_{jg}^{12} + t_{gj}^{21}K_{gj}^{22} - t_{jg}^{21}K_{jg}^{11}} \right)}, \\\nonumber
&&a^{22} = \sum\limits_g {\left( {t_{jg}^{22}K_{jg}^{22} - t_{gj}^{11}K_{gj}^{11}} \right)}, \\\nonumber
&&\bar a^{11} = \sum\limits_g {\left( {\bar t_{jg}^{22}\bar K_{jg}^{22} - \bar t_{gj}^{11}\bar K_{gj}^{11}} \right)}, \\\nonumber
&&\bar a^{12} = \sum\limits_g {\left( {\bar t_{gj}^{11}\bar K_{gj}^{12} - \bar t_{jg}^{22}\bar K_{jg}^{12} + \bar t_{gj}^{21}\bar K_{gj}^{22} - \bar t_{jg}^{21}\bar K_{jg}^{11}} \right)}, \\\nonumber
&&\bar a^{21} = \sum\limits_g {\left( {\bar t_{jg}^{11}\bar K_{jg}^{21} - \bar t_{gj}^{22}\bar K_{gj}^{21} + \bar t_{jg}^{12}\bar K_{jg}^{22} - \bar t_{gj}^{12}\bar K_{gj}^{11}} \right)}, \\\nonumber
&&\bar a^{22} = \sum\limits_g {\left( {\bar t_{gj}^{22}\bar K_{gj}^{22} - \bar t_{jg}^{11}\bar K_{jg}^{11}} \right)}.
\label{eq:kincor_a}
\end{eqnarray}
Definitions of the kinematic correlators are $K_{jg}^{11} = \left\langle {X_j^{\sigma 0}X_g^{0\sigma }} \right\rangle $, $K_{jg}^{12} = \left\langle {X_j^{\sigma 0}X_g^{\bar \sigma S}} \right\rangle $, $K_{jg}^{21} = \left\langle {X_j^{S\bar \sigma }X_g^{0\sigma }} \right\rangle $, $K_{jg}^{22} = \left\langle {X_j^{S\bar \sigma }X_g^{\bar \sigma S}} \right\rangle $, $\bar K_{jg}^{11} = \left\langle {X_j^{\bar \sigma 0}X_g^{0\bar \sigma }} \right\rangle $, $\bar K_{jg}^{12} = \left\langle {X_j^{\bar \sigma 0}X_g^{\sigma S}} \right\rangle $, $\bar K_{jg}^{21} = \left\langle {X_j^{S\sigma }X_g^{0\bar \sigma }} \right\rangle $, $\bar K_{jg}^{22} = \left\langle {X_j^{S\sigma }X_g^{\sigma S}} \right\rangle $. Kinematic correlators are calculated self-consistently with filling numbers of local eigenstates and chemical potential.

\bibliography{refers_enhswave}
\end{document}